%% file: arXiv-Revised-Apr2025.tex
\documentclass[%
prx,twocolumn,
notitlepage,
superscriptaddress,
amsmath,
amssymb,
aps,
prx,
floatfix,
longbibliography
]{revtex4-1}
\usepackage[T1]{fontenc}

\usepackage{natbib}
\usepackage{color}
\usepackage{amsmath}
\usepackage{amssymb}
\usepackage{bm}
\usepackage{graphicx}
\usepackage[usenames,dvipsnames]{xcolor}

\usepackage[%
  colorlinks=true,
  urlcolor=blue,
  linkcolor=blue,
  citecolor=blue
]{hyperref}

\usepackage{array}
\usepackage{ulem}
\usepackage{tikz}
\usetikzlibrary{calc,decorations.markings}

\definecolor{bleu}{rgb}{0.16,0.2.5,0.36}
\definecolor{darkGreen}{RGB}{4,161,85}

\newcommand{\Wload}{W_{\mathrm{load}}} 
\newcommand{\Lload}{L_{\mathrm{load}}} 

\newcommand{\epsbar}{\bar{\epsilon}}
\newcommand{\epsloc}{\epsilon_{\mathrm{loc}}}
\newcommand{\epsplan}{\epsilon_{0}}
\newcommand{\tepsilon}{\tilde{\epsilon}}
\newcommand{\tbar}{\bar{T}}

\newcommand{\EDF}{Extended Data Fig.}

\begin{document}
\pdfoutput=1

\title{Localized tension-induced giant folding in unstructured elastic sheets}

\author{Kexin Guo}\thanks{These authors contributed equally.}
\affiliation{School of Mechanical and Aerospace Engineering, Nanyang Technological University, Singapore 639798, Singapore}

\author{Marc Su\~{n}\'{e}}\thanks{These authors contributed equally.}
\affiliation{Mathematical Institute, University of Oxford, Woodstock Rd, Oxford, OX2 6GG, UK}

\author{Kwok Ming Li}
\affiliation{School of Mechanical and Aerospace Engineering, Nanyang Technological University, Singapore 639798, Singapore}

\author{K. Jimmy Hsia}
\email{kjhsia@ntu.edu.sg}
\affiliation{School of Mechanical and Aerospace Engineering, Nanyang Technological University, Singapore 639798, Singapore}
\affiliation{School of Chemistry, Chemical Engineering and Biotechnology, Nanyang Technological University, Singapore 639798, Singapore}

\author{Mingchao Liu}
\email{m.liu.2@bham.ac.uk}
\affiliation{School of Mechanical and Aerospace Engineering, Nanyang Technological University, Singapore 639798, Singapore}
\affiliation{Department of Mechanical Engineering, University of Birmingham, Birmingham, B15 2TT, UK}

\author{Dominic Vella}%
\email{dominic.vella@maths.ox.ac.uk}
\affiliation{Mathematical Institute, University of Oxford, Woodstock Rd, Oxford, OX2 6GG, UK}


\keywords{ Buckling $|$ Metamaterials $|$ Linear elasticity }

\begin{abstract}
Buckling in compression is the archetype of elastic instability: when compressed along its longest dimension, a thin structure such as a playing card will buckle out-of-plane accommodating the imposed compression without a significant change of length.
However, recent studies have demonstrated that \emph{tension} applied to sheets with microscopic structure leads to out-of-plane deformation in applications from `groovy metasheets' for multi-stable morphing to kirigami grippers. Here, we demonstrate that this counter-intuitive behavior --- a large transverse folding induced by a relatively small imposed longitudinal tension --- occurs also in unstructured sheets of isotropic material. The key to this behavior is that a localized uniaxial tension induces giant folding; we refer to this as `localized TUG folding' to reflect the importance of localized tension and its mode of actuation. We show that localized TUG folding  occurs because of an efficient transfer of applied tensile load into compression --- a geometric consequence of a localized applied tension. We determine scaling results for the folding angle as a function of applied strain in agreement with both experiments and simulations. The generic nature of localized TUG folding suggests that it might be utilized in a broader range of materials and structures than previously realized.
\end{abstract}

\date{\today}

\maketitle

Buckling is  associated with compressive loads on a slender structure \cite{Reis2015}; from railway tracks in extreme heat \cite{Pulcillo2016} to microtubules in cytoplasm \cite{Brangwynne2006}, axial compression is relieved by out-of-plane buckling. Given its ubiquity, it is unsurprising that buckling under compression is  considered a fundamental mechanism through which structures from bridges to marine and aerospace structures may fail \cite{jones2006buckling,Reis2015}. However, other examples have emerged in which an applied tension can lead to a large out-of-plane deformation --- the hallmark of buckling. The first examples of this `buckling in tension' required ingenious design of structures \cite{Zaccaria2011}, but are now also understood to exist in natural systems such as the luxation of human fingers \cite{Fraldi2023}.

\begin{figure*}[t!]
\centering
\includegraphics[width=1.0\linewidth]{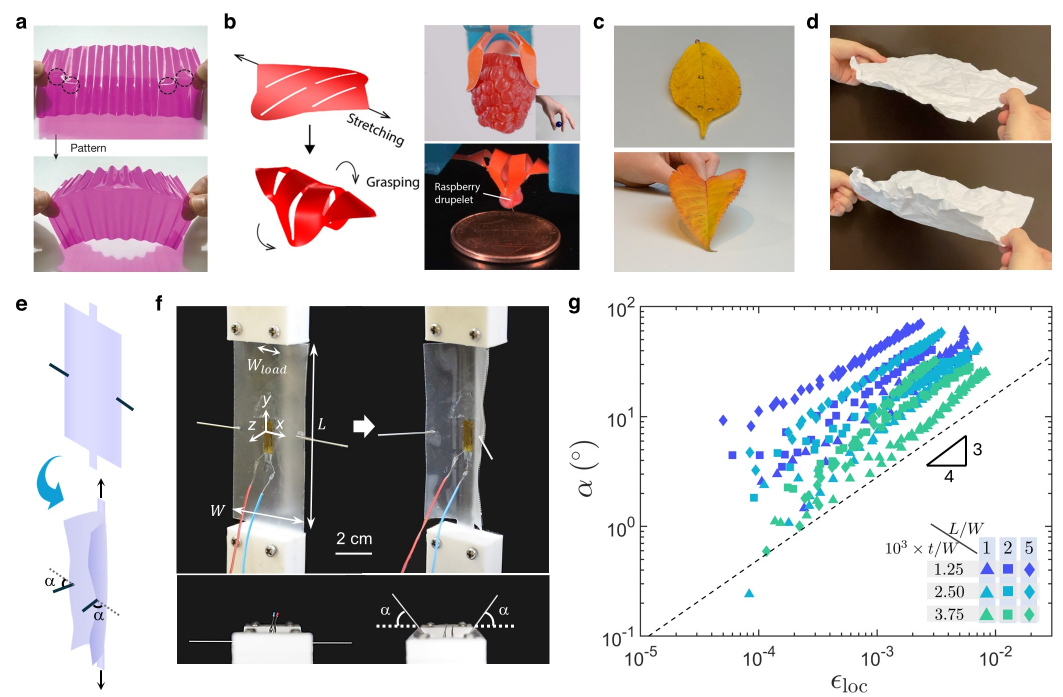}
\caption{\textbf{Uniaxial, localized tension leads to transverse buckling in a range of structured and unstructured thin sheets.} 
\textbf{a} A groovy plastic sheet is stretched by point force along the horizontal direction and forms a scar along which the sheet bends and folds. Adapted from \cite{Meeussen2023}.
\textbf{b} A soft gripper made of a kirigami shell is actuated by stretching along the midline to grasp a raspberry. Adapted from \cite{yang2021grasping}.
\textbf{c} Leaves of the flowering cherry (\emph{Prunus `kanzan'}) show a similar folding at different stages of leaf senescence from early (top) to mid (bottom).
\textbf{d} A crumpled paper sheet is stretched along the horizontal direction by a localized force,  and demonstrates folding in the transverse direction.
\textbf{e} Schematic of the experimental setup. A rectangular  planar sheet is clamped at the additional `tabs' centered on each short side (upper diagram). The sheet folds transversally, developing a non-zero fold angle $\alpha$, as the clamps are displaced to impose a longitudinal displacement (lower diagram).
\textbf{f}~Our experimental setup and associated geometrical parameters of the sheet. Side (upper snapshots) and zenithal view (lower snapshots) of the planar, unloaded, sheet (left), and the loaded sheet when it folds in the transverse direction (right).
\textbf{g}~Experimentally measured inclination angles $\alpha$ as a function of the local tensile strain, $\epsloc$, measured by a strain gauge for a PVC sheet. Plotting on a doubly-logarithmic scale for sheets of various dimensions reveals the power-law relationship $\alpha\propto \epsloc^{3/4}$, which we seek to explain in this article. For all specimens shown here, $W=40\mathrm{~mm}$, $\Wload/W=1/4$, and other geometric parameters are listed in the legend. The Young's modulus is approximately $3\mathrm{~GPa}$ from tensile test measurements on fully clamped strips (see Fig.~\ref{fig:S-ExpSetup}c in the SI Appendix); the Poisson's ratio  $\nu\approx0.38$ for PVC materials. 
}
\label{fig:Examples}
\end{figure*}

At the same time, there has been increased interest in mechanical metamaterials \cite{bertoldi2017flexible,jiao2023mechanical}: elastic objects with a structure that allows for novel functionality. Several examples of these metamaterials seem to buckle in an orthogonal direction under uniaxial extension, see Fig.~\ref{fig:Examples}a,b. For example, groovy metasheets (Fig.~\ref{fig:Examples}a) \cite{Meeussen2023} and kirigami grippers (Fig.~\ref{fig:Examples}b) \cite{yang2021grasping} look superficially different but both exhibit significant transverse motion upon a small axial extension. Similar behavior can be seen in other systems too, including the ribbed sheet \cite{Siefert2021}, while the leaves of some trees fold as senescence progresses (Fig.~\ref{fig:Examples}c).

As a first clue that this behavior may not need the careful design that might first be assumed, the reader is invited to crumple a piece of paper, open it out again, clamp the center between thumb and finger and then pull along one axis. As shown in Fig.~\ref{fig:Examples}d, this informal experiment suggests a similar phenomenology: the paper folds perpendicular to the direction of pulling. A more formal version of this phenomenon has also recently been observed in simulations of (thermal) graphene in which one edge is clamped, leading to extension along this edge \cite{Chen2022,Valenzuela2023}. This ubiquity of the tensile buckling in these systems (both of which are often modelled as isotropic solids, albeit with modified effective moduli) suggests something more fundamental that has not been reported previously. A clue to the key ingredients may be understood by modifying the crumpled paper example just considered: if instead of pulling at the center with two fingers, the crumpled paper is firmly clamped by hand along its whole width and pulled, no folding is observed.

Perhaps the most striking feature of the examples shown in Fig.~\ref{fig:Examples}a-d is that a very small and localized tensile deformation in-plane induces a very large folding angle; for this reason we refer to this means of inducing a fold angle as localized Tension-indUced Giant folding, or localized TUG folding. In this paper, we use a combination of experiments, numerical simulations and mathematical modelling to understand the origin of localized TUG folding. 

 Through the application of the F\"oppl-von K\'arm\'an (FvK) plate theory \cite{Landau1986}, we examine both the onset of  out-of-plane deformation (the near-threshold regime) and its evolution far beyond the critical imposed strain (the far-from-threshold regime~\cite{cerda2002,cerda2003geometry,DavidovitchSchroll11}). We derive the scaling law for the inclination angle as a function of the applied strain that explains the results in experiments and simulations. This scaling law provides useful insight into the geometry and material properties that allow for a large actuation of an elastic sheet.

\section*{Importance of localized tension}
The difference between pulling at a point versus along the full edge motivates only using a localized pulling force. Indeed, similar phenomenology can be observed by taking a uniform sheet and stretching with a small clamped region at the centre (Fig.~\ref{fig:Examples}e). (In the case of uniform clamping along the whole edge, the sheet remains planar --- see the  numerical simulations of a stretched sheet with fully clamped edges in the SI Appendix (Fig.~\ref{fig:S-SimContourPlot}b) --- 
suggesting that this behavior occurs at much smaller strains than are required to cause global wrinkling \cite{cerda2002,cerda2003geometry}.) This phenomenology cannot, therefore, be caused only by the structure, cuts or folds, in the examples already discussed (Fig.~\ref{fig:Examples}a-d). We therefore focus  on experimental realizations in which a small region along the axis is loaded in sheets that are  unstructured (i.e.~spatially homogeneous and isotropic).

In our experiments, we use rectangular sheets of polyvinyl chloride (PVC), with uniform thickness $t$, that are laser cut with length $L$ and width $W$ and an additional `tab' (of width $\Wload=10\mathrm{~mm}$ and length $\Lload=30\mathrm{~mm}$) on each side  to ensure that a localized uniform clamping could be applied. The dimensions of the rectangular sheet (excluding the tab) are varied ensuring that the resulting sheet has $L/W\geq1$ with $W/t\gg1$, see Fig.~\ref{fig:Examples}f. (Further details of the sheets used in experiments may be found in \emph{Methods} and the SI Appendix; images of our experimental setup are shown in Fig.~\ref{fig:S-ExpSetup}.)
The rectangular sheets are clamped by the tabs in 3D-printed fixtures that are attached onto mechanical sliders (Fig.~\ref{fig:Examples}f).
The supports are displaced along the $y$-axis by a distance
\begin{equation}
\Delta u_y\equiv u_y(x=0,y=L/2)-u_y(x=0,y=-L/2)>0.
    \label{eqn:Deltauy}
\end{equation}
 If the sheet remained planar, so that its mid-plane $z=0$ (see Fig.~\ref{fig:Examples}f), this imposed displacement would lead to a planar strain 
 \begin{equation}
\epsplan=\frac{\Delta u_y}{L}.
    \label{eqn:epsplan}
 \end{equation}
 However, once the strain reaches a small threshold value of  it curls in both $x$ and $y$ directions so that the mid-plane of the sheet is deformed to $z=\zeta(x,y)$. Double curving is displayed by a projected laser  line along the two midlines at $x=0$ and at $y=0$ --- images are included in Fig.~\ref{fig:S-ExpSetup}e,f of the SI Appendix --- and distinguishes this effect from the edge clamping-induced tilting of graphene \cite{Chen2022,Valenzuela2023}, in which the imposed boundary condition inhibits this double curving. Figures~\ref{fig:Examples}e,f show the transverse curving as viewed with small cannula needles attached to the edge of the sheet; these needles allow us to measure the angle, $\alpha$, that the edge of the sheet makes with the horizontal.

\section*{Scaling behavior of folding}
The angle, $\alpha$, that the edge of the sheet makes with the horizontal is measured as a function of imposed strain, see Fig.~\ref{fig:Examples}g. (Here the value of the strain is measured using a strain gauge placed at the center of the sample, which gives a measure of the local strain $\epsloc$ --- see \emph{Methods}.) The experimental results  of Fig.~\ref{fig:Examples}g suggests that $\alpha\propto\epsloc^{3/4}$
--- a scaling that is also consistent with similar experiments on crumpled paper (see Fig.~\ref{fig:CrumpledSheets} of the SI). This  scaling  is novel in two important respects: firstly, we reiterate that $\epsloc>0$ --- the sheet is subject to a \emph{tensile} load and yet appears to buckle in the transverse direction. Secondly, this scaling is more sensitive to the imposed end-stretching than the regular (compressive) buckling in which, for example, the amplitude $A\propto \Delta u_y^{1/2}$~\cite{Howell2008} where $\Delta u_y$ is the imposed end-shortening defined in \eqref{eqn:Deltauy}.

\begin{figure*}[t!]
\centering
\includegraphics[width=1.0\linewidth]{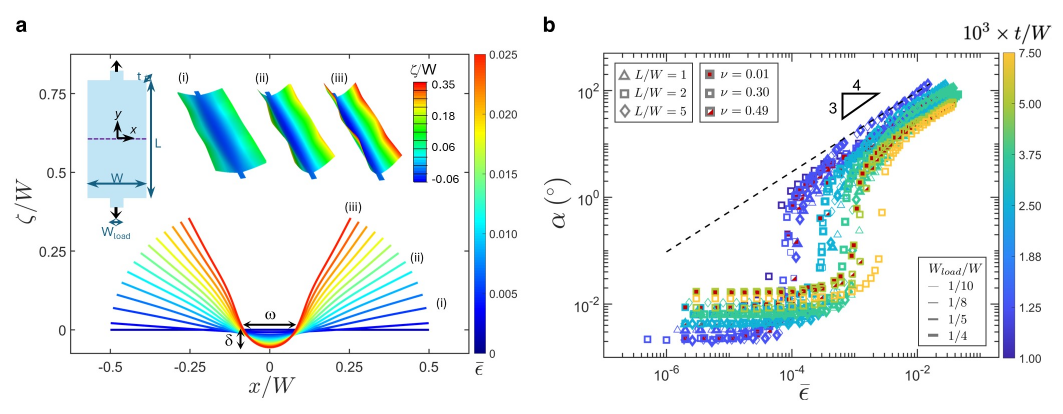}
\caption{\textbf{Finite element simulations of tensile stretching induced folding in elastic sheets.} 
\textbf{a} Cross-sectional  profiles, $\zeta(x,0)$, of a  sheet stretched along the $y$-axis  in normalized units. The color bar indicates the tensile strain of the midline at $x=0$ (see top-left inset). The top-right insets (i-iii) show the deformed shapes of the elastic sheet at increasing strains, with corresponding numbers labelled in the cross-sectional profiles, and colors indicating the $\zeta$-displacements normalized by $W$ according to the color bar. The model shown here has the following geometric and material parameters: $L/W=2, t/W=2.5\times 10^{-3}, \Wload/W=1/10$, $E=3\mathrm{~GPa}$, and $\nu=0.3$.
\textbf{b} Simulation results of tensile strains and inclination angles for elastic sheets with various geometric parameters and Poisson's ratio, as indicated in the legend. The color bar indicates the thickness-to-width ($t/W$) ratio. These numerical results suggest that $\alpha\propto \epsbar^{3/4}$ (dashed line), as also observed experimentally. Note that in (b) the marker style is used to encode the geometry and Poisson's ratio used in simulations (as indicated by the grayscale legends). Color is used to encode the thickness (as per the colorbar on the right hand side).
 }
\label{fig:FEMSimulation}
\end{figure*}

To test whether this behavior is indeed generic, or is caused by some imperfection or anisotropy of the experimental system, we also performed systematic numerical simulations using the finite element method (FEM) software ABAQUS. The results of our FEM simulations allow the deformed shape $\zeta(x,y)$ to be calculated. In particular, the observed inclination angle is plotted as a function of the mean strain
\begin{equation}
    \epsbar=\epsplan+\frac{1}{2L}\int_{-L/2}^{L/2}\left(\frac{\partial \zeta}{\partial y}\right)^2~\mathrm{d}y,
    \label{eqn:EpsBarDefn}
\end{equation}  in Fig.~\ref{fig:FEMSimulation}b. These results demonstrate that the key phenomenology is similar: out-of-plane bending occurs transverse to the imposed stretching (see Fig.~\ref{fig:FEMSimulation}a, which displays $\zeta(x,0)$, the vertical displacement along the midline at $y=0$), while the angle of inclination, $\alpha$, follows a power-law similar to that observed experimentally, cf.~Fig.~\ref{fig:Examples}g. Numerical simulations allow us to test the role of the Poisson's ratio $\nu$; our results show a small effect of $\nu$ but, crucially, that the effect persists as $\nu\to0$ (see Fig.~\ref{fig:FEMSimulation}b). This is in contrast with the tensional wrinkling experiment of Cerda and Mahadevan (CM) \cite{cerda2003geometry} for which the effect disappears entirely as $\nu\to0$ \cite{XinDavidovitch21,XinDavidovitch21-2}. We further discuss the role of Poisson's effect in the subsequent theoretical analysis.

One important difference between experiments and these numerical simulations is that we observe a small (experimentally-undetectable), and $\epsbar$-independent, angle for very small stretching with a sharp transition to a larger, $\epsbar$-dependent, $\alpha$, at some critical strain, $\epsbar_c$. Note that  this transition is not observable in experiments because of the limited angular resolution, which is caused by small fluctuations in the  initial state. We therefore seek to understand both the emergence of a critical strain, $\epsbar_c$, and the scaling of the angle $\alpha$ with $\epsbar>\epsbar_c$.

Our experiments and numerical simulations show a wide range of inclination angles $0.01^\circ\lesssim\alpha\lesssim100^\circ$ at small imposed strains --- the axial strain is at most $\epsbar\sim10^{-2}$ in Figs.~\ref{fig:Examples} and \ref{fig:FEMSimulation}.
The conjunction of large out-of-plane deflections and small strains is a common feature of thin sheets and is encapsulated in the FvK plate theory, which combines a mechanically linear response (i.e.~a Hookean stress-strain relationship) with nonlinear displacement-strain relationships (i.e.~out-of-plane displacements contribute to strains) \cite{Landau1986}. The coupling of the stress field within the sheet to the sheet's shape in the FvK equations makes analytical progress difficult. A key simplification comes  while the sheet remains planar: we then need only solve for the stress within the loaded sheet and the characteristic tension is linear in the imposed  end separation,
\begin{equation}
T_0=Y\epsplan,
\label{eqn:T}
\end{equation}where $Y=E\,t$ is the stretching modulus; $E$ denotes the Young’s modulus, which is measured experimentally (see SI), but does not directly control this geometrical effect. The problem for the planar stress can be posed as the bi-harmonic equation for the Airy stress function $\chi$ with appropriate boundary conditions (see the mathematical model in the SI Appendix). The resulting equations can be analyzed using Laplace transform methods \cite{Williams52,Benthem63,MorseFeshbachI} (see SI Appendix for details). The key result of this analysis is that, for sufficiently slender sheets at least, and for $\Wload\ll W$, the stress components within the sheet remain tensile \emph{except} within a small region close to the edges (i.e.~within a typical distance $W$ from the short edges the transverse stress is compressive) (see Fig.~\ref{fig:PlanarStressState}). Moreover, this analysis allows us to compare the effect of point-loading, $\Wload\ll W$, with the more common full-width loading, $\Wload=W$, in the limit of vanishing Poisson ratio $\nu\to0$. This comparison (see Fig.~\ref{fig:PlanarStressTransverse} of the SI Appendix) shows that the zone of compression persists with point-loading but disappears with full-width loading --- the geometry of the loading is key to the localized TUG-folding demonstrated here and distinguishes it from the tensional wrinkling studied extensively previously \cite{cerda2003geometry}. In addition, this effect does \emph{not} require a negative Poisson ratio, as might be inferred on the basis of the tilt phase in clamped thermal graphene \cite{Chen2022}. The origin of this compression signals the importance of a `mechanism' in generating the compression (see Fig.~\ref{fig:PlanarStressTransverse} of the SI Appendix).

The linearity of this planar problem for the stress prior to buckling, means that the magnitude of the compressive stress generated by `tugging' must be linear in the characteristic tension $T_0$.
The coincidence of lengths (i.e.~that the lateral scale of the compressed area scales with $W$) and of stresses (i.e.~that the resulting transverse compression is linear in the characteristic tension, $T_0$) simplifies the near-threshold analysis of instability~\cite{XinDavidovitch21} so that the classic Euler-like buckling analysis is helpful (see the mathematical model in the SI Appendix for details). Hence, we expect that buckling will occur when $T_0$ reaches some multiple of $B/W^2$, which immediately gives us that
\begin{equation}
    \epsplan^c=\epsbar_c\propto \frac{B}{YW^2}.
    \label{eqn:EpsCrit1}
\end{equation} For the unstructured, isotropic sheets used here, the bending modulus $B=Yt^2/[12(1-\nu^2)]$ and so \eqref{eqn:EpsCrit1} predicts that 
\begin{equation}
    \epsplan^c=\epsbar_c\propto \frac{t^2}{(1-\nu^2)W^2},
    \label{eqn:EpsCrit}
\end{equation} which may be interpreted as the reciprocal of the von K\'{a}rm\'{a}n number \cite{Kosmrlj2017}.  We emphasize that the scalings for the critical strain in \eqref{eqn:EpsCrit1} and \eqref{eqn:EpsCrit} are the standard scalings for Euler buckling \cite{Lidmar2003,Chen2022} --- we shall see later that they are borne out by detailed numerical simulations, which serve to show that this transition is indeed driven by buckling. 

We also observe that the sheet is doubly curved in a narrow region (see Fig.~\ref{fig:FEMSimulation}a); outside this region, there is a small curvature in the $y$-direction (along the pulling direction), while in the $x$-direction, the sheet is essentially flat (albeit inclined at the angle $\alpha$) --- in this sense the response of the sheet to uniaxial tension is to fold in the transverse direction. We denote the (unknown) width of this region by $\omega$ (see Fig.~\ref{fig:FEMSimulation}a) and estimate it using the experimental (and numerical) observation of the shape to pose an ansatz for the sheet's out-of-plane displacement, $\zeta(x,y)$, relative to the initially planar state; in particular, we take:
\begin{equation}
    \zeta(x,y)=\delta\times\begin{cases}
\cos\frac{\pi y}{L}\cos\frac{\pi x}{\omega},\quad |x|<\omega/2,\\
    \frac{\pi}{\omega}\left(\frac{\omega}{2}-|x|\right),\quad \omega/2\leq |x|\leq W/2,
        \end{cases}
    \label{eqn:ansatz}
\end{equation} 
where $\delta$ is the maximum amplitude (see Fig.~\ref{fig:FEMSimulation}a), and the constants have been chosen to ensure that the vertical displacement, slope and curvature of the sheet are all continuous at the edge of the curved central region. (Note that this ansatz is only appropriate for $\Wload\ll\omega$, since otherwise a third region would be required.) To determine $\omega$, we note that the smaller its value, the more curved the sheet is in the central region, and the higher its bending energy is: $U_{\mathrm{bend}}\sim B\int~(\nabla^2\zeta)^2\mathrm{d}A\sim B\delta^2 L/\omega^3$ (where we have neglected the bending energy caused by the curvature in the $y$-direction because $L\gtrsim W\gg\omega$). Conversely, the excess stretching energy caused by being doubly-curved is localized to this central region so that $U_{\mathrm{stretch}}\sim\int\sigma:\epsilon~\mathrm{d}A\sim WL\tbar^2/Y+\omega L \tbar\delta^2/L^2$, where the characteristic tension, $\tbar$, is linear in the mean strain,
\begin{equation}
\tbar=Y\epsbar,
\label{eqn:tbar}
\end{equation} and we make the common assumption that the sheet buckles in the compressing ($x$) direction to relax the compressive stress. Notice also that there are two different length scales for the two terms in $U_{\mathrm{stretch}}$: the whole sheet, of width $W$, feels the longitudinal tension, but only the narrow strip, of width $\omega$, is subject to the additional strain, $\sim \delta^2/L^2$, from double-curvature. In the first term in $U_{\mathrm{stretch}}$ we also assume that $\epsplan\sim \epsbar=\tbar/Y$ remains true after buckling, by virtue of the observation that $\epsbar-\epsplan\propto \epsplan-\epsbar_c$ sufficiently close to the  buckling threshold ($(\epsbar-\epsbar_c)/\epsbar_c\lesssim 10$ --- see the numerical simulation results for the mean strain beyond buckling in Fig.~\ref{fig:S-DeltayEpsbar}b in the SI Appendix). The total elastic energy, $U_{\mathrm{bend}}+U_{\mathrm{stretch}}$, is then minimized when 
\begin{equation}
    \omega\propto \left(\frac{BL^2}{\tbar}\right)^{1/4}=\left(\frac{BL^2}{Y\epsbar}\right)^{1/4}.
    \label{eqn:CurvedWidth}
\end{equation} Note that the scaling law for the width of the curved region, \eqref{eqn:CurvedWidth}, is identical to the scaling for the wavelength of wrinkling observed when a sheet has its whole width clamped and is then pulled perpendicular to this width, as studied by CM \cite{cerda2003geometry}. We will discuss the similarities and the differences between our results and CM later in the text.

Geometry gives that the inclination angle satisfies $\alpha\approx\tan\alpha=\pi\delta/\omega$ in which $\omega$ has been determined (at least to a scaling level) by the previous argument and we assume $\alpha\ll1$ for consistency with our use of the FvK formalism; the final part of the puzzle therefore  is to determine the maximum amplitude $\delta$ in \eqref{eqn:ansatz}. 
We note that  our numerics show that, even after onset, $\epsbar\propto \epsplan$ (see Fig.~\ref{fig:S-DeltayEpsbar}b of the SI Appendix). As a result, we have that $\delta^2/L^2\propto\epsbar-\epsbar_c$ and  $\delta\propto L(\epsbar-\epsbar_c)^{1/2}$, which, when combined with the geometrical relationship $\alpha\approx\pi\delta/\omega$, gives us that
\begin{equation}
    \alpha\propto \delta \left(\frac{Y\epsbar}{BL^2}\right)^{1/4}\propto \left(\frac{YL^2}{B}\right)^{1/4}\epsbar^{1/4}(\epsbar-\epsbar_c)^{1/2}.
    \label{eqn:InclinationAngle}
\end{equation} Using \eqref{eqn:EpsCrit}, this result may also be written
\begin{equation}
   \alpha\propto \frac{t L^{1/2}}{W^{3/2}\sqrt{1-\nu^2}}\left(\frac{\epsbar}{\epsbar_c}\right)^{1/4}\left(\frac{\epsbar}{\epsbar_c}-1\right)^{1/2}.
    \label{eqn:InclinationAngle2}
\end{equation}
Here, we have employed an empirical observation, based on our numerics, to understand the scaling of the amplitude $\delta$ with strain. Beyond threshold, the non-linear coupling between strain and vertical displacement hinders analytical progress such that even approximate theories based on the FvK framework, e.g.~tension field theory (see e.g.~the work of Xin and Davidovitch~\cite{XinDavidovitch21-2} on the CM problem~\cite{cerda2003geometry}), ultimately require numerical resolution.

\begin{figure}[h]
\centering
\includegraphics[width=1.0\linewidth]{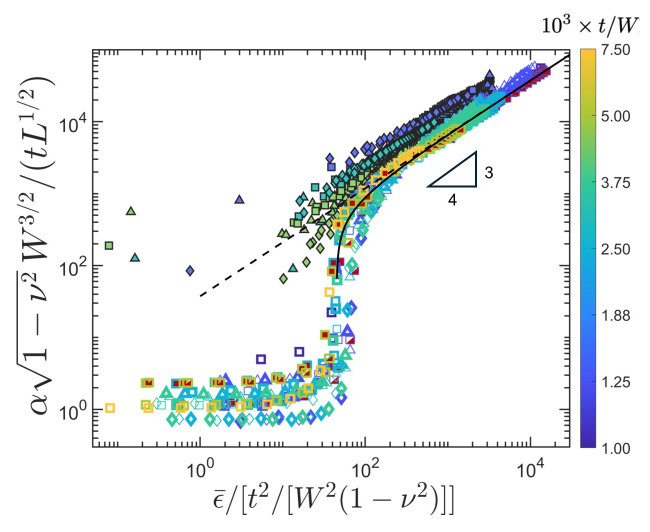}
\caption{\textbf{Master curve for inclination angle,  $\alpha$ ($^{\circ}$), as a function of mean strain, $\epsbar$.} Here, results are shown for both simulations and experiments with rescalings suggested by the critical strain at buckling, \eqref{eqn:EpsCrit}, and the scaling for inclination angle \eqref{eqn:InclinationAngle2}. The dashed line shows the expected scaling behavior for $\epsbar\gg\epsbar_c$ while the solid curve shows the result \eqref{eqn:InclinationAngle2} assuming a prefactor of $43$ in \eqref{eqn:EpsCrit}. (The same legends as in Fig.~\ref{fig:FEMSimulation}b are adopted for simulation data, i.e.~marker shapes denote the aspect ratio, marker fill indicates the Poisson's ratio, marker line-widths denote $\Wload/W$ and color indicates the sheet thickness as per the color bar.) Experimental data points are shown behind the simulation data in black-bordered markers with fill colors corresponding to the sheet thickness and shapes denoting the aspect ratio. (The experimentally measured strain $\epsloc$ has been converted to an effective mean strain $\epsbar_{\mathrm{eff}}$ based on simulation results --- see the plot of $\epsloc$ vs $\epsbar$ in Fig.~\ref{fig:S-LocGlob} in the SI Appendix. Moreover, to avoid cluttering the plot, prior to buckling every second data point from each simulation is displayed while in the post-buckling regim every eighth data point is displayed.)} 
\label{fig:ExpEpsBar}
\end{figure}

To highlight the trend that $\alpha\to0$  as $\epsbar\searrow\epsbar_c$, we rescale the $x$-axis in Fig.~\ref{fig:ExpEpsBar} on the basis of the predicted scaling for $\epsbar_c$ given in \eqref{eqn:EpsCrit}; \eqref{eqn:InclinationAngle}, in turn, suggests a rescaling for the inclination angle on the $y$-axis. The results plotted in this way show a reasonable collapse (see Fig.~\ref{fig:ExpEpsBar}). The experimental data points in this figure are plotted using an effective mean strain obtained by the linear relation in Fig.~\ref{fig:S-LocGlob}a to keep a consistent strain definition with the numerics; a similar collapse is observed when plotting $\epsloc$ in the experimental results (see the collapse of the inclination angle in Fig.~\ref{fig:S-LocGlob}b in the SI Appendix).

\begin{figure*}[t!]
\centering
\includegraphics[width=0.95\linewidth]{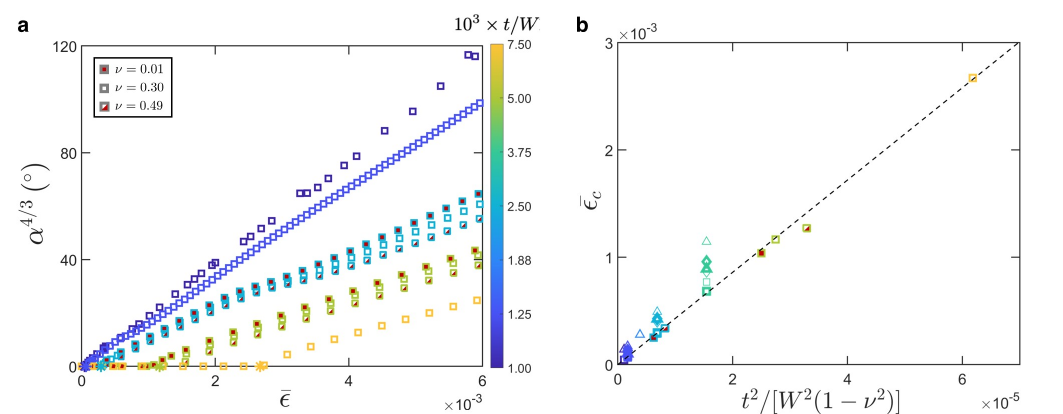}
\caption{\textbf{Critical buckling behavior in FEM simulations.} 
\textbf{a} The inclination angle $\alpha$ allows for a clearer determination of the critical strain for buckling in our FEM simulations.  Buckling occurs where the linear region of the scaled inclination angle intersects with the $x$-axis, which defines the critical buckling strain $\epsbar_c$. The curves shown all have $L/W=2$, $\Wload/W=1/5$, with various sheet thickness ratios as indicated by the color bar. 
\textbf{b} The critical strains for various geometric parameters and Poisson's ratios $\nu$ are proportional to the ratio $B/(YW^2)\propto (t/W)^2/(1-\nu^2)$; the dashed line shows the relationship \eqref{eqn:EpsCrit} with coefficient of proportionality $43$.}
\label{fig:Threshold}
\end{figure*}

\subsection*{Critical strain for buckling}
The prediction for the inclination angle also gives us a more precise means to determine the threshold strain for buckling in our FEM simulations: except very close to the buckling transition, we expect $\alpha\propto(\epsbar-\epsbar_c)^{3/4}$ and so plot $\alpha^{4/3}$ as a function of $\epsbar$ (see Fig.~\ref{fig:Threshold}a). This plot shows the expected linear behavior and also allows us to determine the threshold strain, $\epsbar_c$, as a function of the sheet thickness ratio, $t/W$, and Poisson's ratio $\nu$. The resulting plot (Fig.~\ref{fig:Threshold}b) shows that the prediction of \eqref{eqn:EpsCrit} is borne out in our numerical simulations, although points with $L=W$ do not collapse as well as those points with $L/W\gg1$. We also note that extrapolating this critical strain to the experimentally relevant values of $t/W$ suggests the typical onset strain in experiments is around $10^{-4}$, which for a sample with $L=10\mathrm{~cm}$, corresponds to a displacement $\Delta u_y\approx10\mathrm{~\mu m}$. This, in addition to the very small angles close to onset already noted, explains why this threshold behavior is not visible in our experiments.

Comparing the critical buckling behavior in our problem (Fig.~\ref{fig:Threshold}b) and in models of the CM problem (e.g.~Fig.~7A of~\cite{XinDavidovitch21}), we note that both have identical scaling behavior with thickness and width of the sheet. However, the effect described here persists as the Poisson effect vanishes, $\nu\to0$, in contrast to the CM problem. Moreover, we  find that the threshold value of the dimensionless control parameter $\epsbar_c$ is smaller than that in the CM problem \cite{XinDavidovitch21}; for example, comparing the critical strains with $\nu=0.32$ and $t/W=10^{-4}$ we have 
\begin{equation}
    \frac{T_{0}^{c}}{Y} \approx \frac{t^2}{W^2}\times \begin{cases}
43/(1-\nu^2),\quad \qquad \Wload\ll W,\\
        9\times10^4 \quad \Wload=W.
    \end{cases}
    \label{eq-critical-strain-comp}
\end{equation} (Here, the critical value for the full-width case, $\Wload=W$, comes from Fig.~7A of \cite{XinDavidovitch21}, while the critical value for $\Wload\ll W$ comes from our numerical results in Fig.~\ref{fig:Threshold}b.)
Note that with the load applied across the full width, the threshold strain required to trigger buckling is more than $1000$ times larger than in the localized case considered here. To understand this apparently enhanced resistance to buckling in the CM problem, we compare the maximum compressive stress in the full-width setup (as calculated in ref.~\cite{XinDavidovitch21}) with that in our problem (see the details of the mathematical model in the SI Appendix). For the specific case of $L/W=4$ and $\nu=0.3$, we find (see Fig.~\ref{fig:PlanarStressTransverse} of the SI Appendix) that
\begin{equation}
    \max_{y}\left\{\frac{-\sigma_{xx}}{T_{0}}\right\}\approx \begin{cases}
        0.6,\quad \qquad \Wload\ll W,\\
        0.006, \qquad \Wload=W.
    \end{cases}
\end{equation} 
The magnitude of the maximum compressive stress is hence $100$ times larger in the localized case considered here than in the full-width case. Recalling  that the critical strain was $1000$ times larger for the full-width case, we deduce that much of the difference in critical imposed strain can be attributed to the increased efficiency of focusing the longitudinal tensile strain into a transverse compression when pulling at a localized load than with full-width. However, there also remains an effect of the boundary conditions on the relevant  Euler buckling load --- indeed, even a standard buckling analysis gives an increase of a factor of 4 between the critical load for a simply-supported beam (analogous to the freer edges here) and a clamped beam (analogous to the more constrained edges of the CM problem) \cite{Landau1986}.

We also found that the scaling for the width of the curved region~\eqref{eqn:CurvedWidth} is equivalent to the wavelength of wrinkles in CM~\cite{cerda2003geometry}. While we emphasize that, for the narrow loading $\Wload\ll\omega$ we studied here, only a single depression is formed (equivalent to half a wavelength), this equivalence suggests that the narrow curved strip is effectively wrinkling around the narrowly loaded part. If we relax the assumption $\Wload/\omega\ll 1$ --- by imposing longitudinal strains large enough that $\omega\sim\Wload$, for example --- we observe  wrinkles in the curved part of the sheet (see Fig.~\ref{fig:wrinkles} of the SI Appendix). However, the localized-loaded sheet and the fully-clamped sheet have significant differences in that localized TUG folding  persists as $\nu\to0$ and the sheet must be loaded to large strains to observe full-width wrinkling.

\section*{Conclusions}
We have shown that localized \emph{tensile} loading of an unstructured elastic sheet along one axis actuates folding in an orthogonal direction driven by inducing \emph{compression}. A particularly striking feature of this transformation from tension to compression is that it allows a very small and localized strain to actuate a giant folding angle: what we call localized `TUG folding'. We claim that a similar mechanism has been observed, if not directly reported, in structured materials---such as a crumpled sheet of paper (Fig.~\ref{fig:Examples}d), kirigami shells~\cite{yang2021grasping}, and groovy plastic sheets~\cite{Meeussen2023}---but to our knowledge has not been reported in homogeneous, isotropic materials. While our scaling laws for the critical strain and inclination angle reveal how the geometry of the sheet and the properties of the material should be tuned to actuate a homogeneous sheet, in the same way, they also give new insights into the mechanisms by which the various metasheets (in the examples mentioned above) are able to achieve localized TUG folding. We note that these all involve structures (albeit different in nature) that make stretching in one direction easier but making bending in the transverse direction more difficult. For example, the ridges of a crumpled piece of paper buffer excess length \cite{Vella2019} (reducing the effective $Y$), while also increasing the effective thickness (and hence increasing the effective $B$ \cite{Kosmrlj2016}), and similarly for the perpendicular cuts and grooves in kirigami grippers and groovy metasheets, respectively. In an unstructured sheet, the combination of a decreased stretching modulus and an increased bending modulus results in an increased buckling strain --- see ~\eqref{eqn:EpsCrit1}. However, at a given multiple of the strain threshold, $\tepsilon=\epsbar/\epsbar_c$, this increase in $B/(YW^2)$ becomes helpful in achieving localized TUG folding; in particular, we note that \eqref{eqn:InclinationAngle} may be rewritten:
\begin{equation}
    \alpha\propto \left(\frac{B}{YW^2}\right)^{1/2}\left(\frac{L}{W}\right)^{1/2}\tepsilon^{1/4}(\tepsilon-1)^{1/2},
    \label{eqn:FoldScaleGeneral}
\end{equation} and so, for a given $\tepsilon$, large $B/(YW^2)$, combined with $L/W\gg1$, leads to a large folding angle~\footnote{Note that \eqref{eqn:FoldScaleGeneral} is similar to the result \eqref{eqn:InclinationAngle2} except that for more general, possibly structured, solids, we do \emph{not} assume that $B\propto Yt^2$.}. A second key ingredient therefore is the geometrical focusing that reduces the critical $\epsbar_c$ by a factor of $\sim10^3$ compared to the fully-clamped case~\eqref{eq-critical-strain-comp}. The efficiency of this conversion from tension to transverse compression means that a given strain leads to a large $\alpha$; in addition, if $Y$ is decreased by  structure, such as  cuts, or by the microscopic `buffering'  of area \cite{Vella2019}, such as crumples or excess area \cite{Chen2022,Valenzuela2023},  then a fixed load will lead to a still larger strain and hence even larger folding angle. This fundamental insight opens the door for designing structures that can more readily demonstrate localized TUG folding with, or without, structure. It may also lead to new insights into similar phenomenology in systems as diverse as dimpled sheets (which fold upon `snapping' of the dimples along a line \cite{Liu2023}) to tension-induced buckling in active systems  \cite{FierlingJohn22}.

\emph{Acknowledgments}\\
This work was partially supported by the UK Engineering and Physical Sciences Research Council via Grant No.~EP/Y027949/1 (MS and DV); the Ministry of Education, Singapore, via MOE AcRF Tier 3 Award MOE-MOET32022-0002 (KJH); the Presidential Postdoctoral Fellowship from Nanyang Technological University, Singapore, and the start-up funding from the University of Birmingham, UK (ML). We thank Ms Mariona Caus for assisting in taking the pictures in Fig.~\ref{fig:Examples}d.

\emph{Materials and Methods} \paragraph{Sample preparation}
Test specimens were prepared from commercial PVC films with thickness $t\in\{40,100,150\}\mathrm{~\mu m}$. Rectangular sheets were laser cut with width $W=40\mathrm{~mm}$, length $L\in\{40,80,200\}\mathrm{~mm}$, and two tabs with width $\Wload=10\mathrm{~mm}$ and length $\Lload=30\mathrm{~mm}$ at each end (see Fig.~\ref{fig:S-ExpSetup}b of SI Appendix). The fixtures were then attached to the blocks of two linear-stage motors that were controlled manually. The specimens were clamped at the tabs in 3D-printed clamp fixtures designed with teeth on the inner surfaces to reduce slipping. Prior to clamping the specimen, strain gauges with a gauge length of $6 \mathrm{~mm}$, resistance of $350~\Omega$, and resolution of $1~ \mu \epsilon$ (Tokyo Measuring Instruments Laboratory, GFLA-6-350-70)  were adhered to the centre of the specimen using strain gauge adhesives (Tokyo Measuring Instruments Laboratory). The measured strain data were recorded by a data logger (Tokyo Measuring Instruments Laboratory, TDS-303). Inclination angles of the sheets were measured from photographs taken from the orthogonal plane (see the experimental methods in the SI Appendix for further details) with the assistance of cannula needles adhered to the edges of the sheet. The needle lines were detected using MATLAB scripts; the inclination angles reported are the mean values of the left and right sides.

\paragraph{Numerical simulations}
Numerical simulations were conducted using the commercial finite element analysis software ABAQUS/2023. The solid was represented by S4R shell elements and a linear elastic constitutive relation was applied with Young’s modulus $E=3000 \mathrm{~MPa}$; most simulations were performed with Poisson’s ratio $\nu=0.3$, though results with $\nu=0.01$ and $\nu=0.49$ are also reported. The boundary conditions were set as in the experiment: the top edge of the upper tab was fully fixed, while displacement loading was applied along the $y$-axis to the bottom edge of the lower tab. Two analysis steps were conducted in sequence: first, a small out-of-plane displacement were applied to the midline along $x=0$ with a magnitude of $\sim$ 1\% of sheet thickness, introducing initial imperfections to the sheet; second, the static analysis step was performed for the process of tensile stretching and sheet folding.


%

\clearpage
\setcounter{section}{0}
\renewcommand{\thesection}{S\arabic{section}}%
\setcounter{equation}{0}
\renewcommand{\theequation}{S\arabic{equation}}%
\setcounter{figure}{0}
\renewcommand{\thefigure}{S\arabic{figure}}%
\renewcommand{\figurename}{\EDF}%

\onecolumngrid
\begin{appendix}

\renewcommand{\thefigure}{S\arabic{figure}}
\renewcommand{\theHfigure}{S\arabic{figure}}
\setcounter{figure}{0}

\input{SI-Revised.tex}

\end{appendix}

\end{document}

%% file: SI-Revised.tex
%
\newpage

\title{Supplementary Information}
\maketitle

\section{Experimental methods} \label{sec:exp-method}
\begin{figure*}[h]
    \centering
    \includegraphics[width=0.85\linewidth]{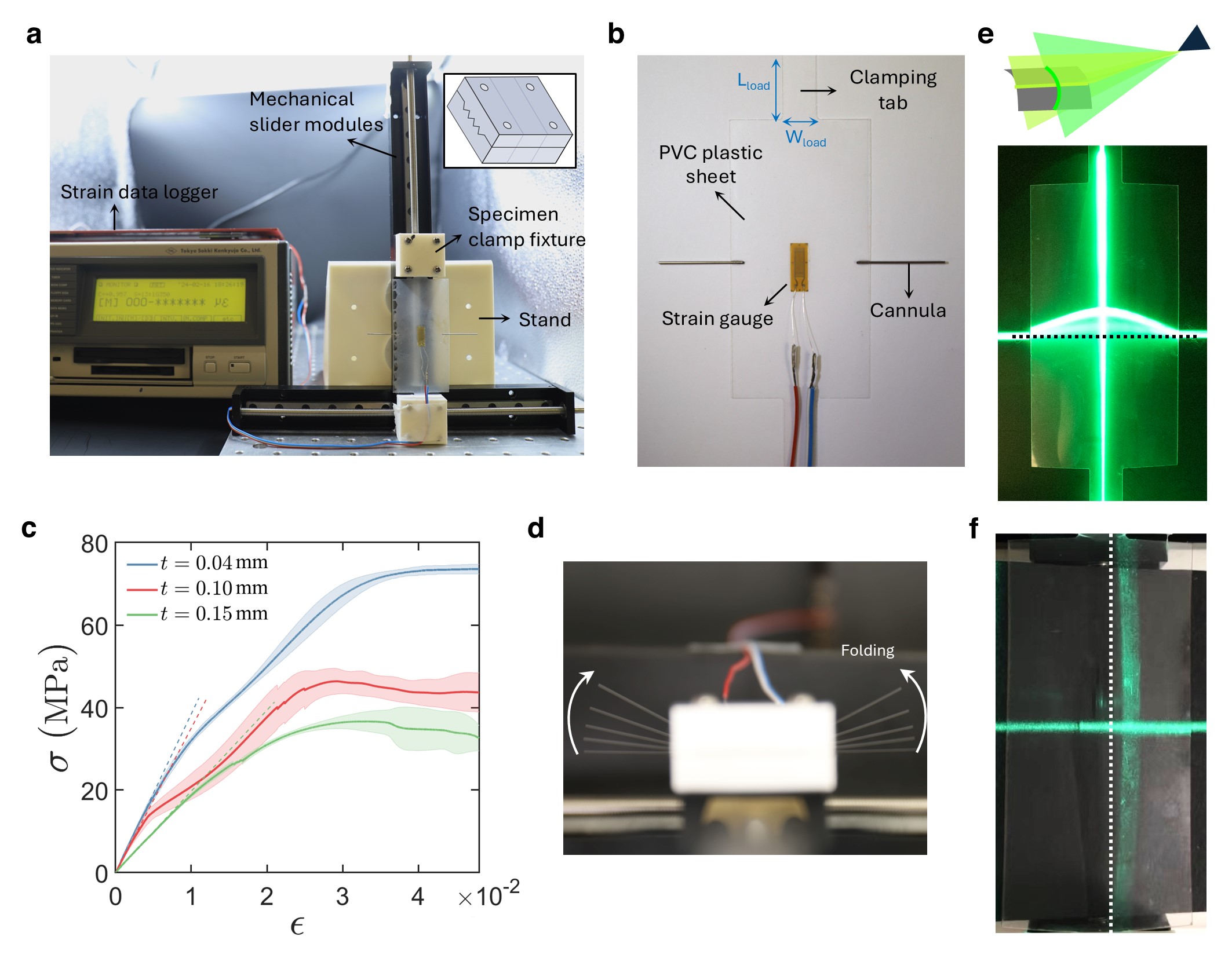}
    \caption{\textbf{Experimental setups.}
    \textbf{a} Tensile-stretching equipment consisting of linear-stage motors, 3D-printed specimen clamp fixtures and a stand (all labelled on the figure).
    \textbf{b} Testing specimen, showing the location of the strain gauge, cannula and where the sheet is clamped (see labels).
    \textbf{c} Stress-strain curves obtained from PVC sheets with fully clamped loading edge, $\Wload=W$; since these tests are performed in a tensile testing machine, the strain $\epsilon$ corresponds to $\epsplan$ used in the main text.
    \textbf{d} Composite image showing the evolution of the folding angle, $\alpha$, with increasing strain.
    \textbf{e} Visualization of the transverse curvature upon stretching via a projected laser line. Schematics illustrate the projection setup. The dotted line marks the straight line from which the beam deviates due to the curvature.
\textbf{f} Visualization of the longitudinal curvature via a projected laser line.}
    \label{fig:S-ExpSetup}
\end{figure*}

    The stretching equipment is set up by two mechanical slider modules with $300\mathrm{~mm}$ range of motion. They are fixed to a stand and a clamping fixture is attached to each module (Fig.~\ref{fig:S-ExpSetup}a). Both the fixtures and the stand were 3D printed (Raise3D FDM Printer) using polylactic acid (PLA). The specimen fixtures are customly designed to fit the slider and the clamp surfaces have teeth-like patterns to enhance grip (Fig.~\ref{fig:S-ExpSetup}a,~inset). The specimens (Fig.~\ref{fig:S-ExpSetup}b) are made of laser-cut commercial polyvinyl chloride (PVC) sheets. 
    
    The elastic modulus of the sheets are determined from the tensile tests of rectangular strips with $\Wload = W$. The strips have, approximately, dimensions $W=20~\mathrm{mm},L=170~\mathrm{mm}$. Averaging five samples for each thickness, we obtain the Young's modulus of PVC sheets with thickness $0.04,0.10,0.15\mathrm{~mm}$ to be $3.84,3.50,1.96\mathrm{~GPa}$, respectively (Fig.~\ref{fig:S-ExpSetup}c). Note also that the  value of the Young's modulus only scales the forces required to cause a given deformation, not the deformation itself. The Poisson's ratio of PVC sheets is 0.38.

    The sheets used in the experiment have extended `tabs' which are regions for clamping with $\Wload=10\mathrm{~mm}$ and $\Lload=30\mathrm{~mm}$ (Fig.~\ref{fig:S-ExpSetup}b). The connection between the clamping regions and the main part of the sheets is designed with rounded corners of radius $1\mathrm{~mm}$ to prevent failure due to stress concentration. Tensile strains in the elastic sheets are measured by strain gauges (Tokyo Measuring Instruments Laboratory, GFLA-6-350-70) with gauge length of $6\mathrm{~mm}$, resistance of $350~\Omega$, and resolution of $1\mathrm{~\mu \epsilon}$. The strain gauges are attached to the centre of the specimen along the loading direction using an adhesive, and the gauge leads were connected to electrical wires that input to the strain data logger (Tokyo Measuring Instruments Laboratory, TDS-303). When clamping specimens between fixtures, vertical alignment is ensured for symmetric loading. Cannula needles with an outer diameter of $700\mathrm{~\mu m}$ are adhered to the $y$-middle of the sheet near the edges as an indicator in the measurement of inclination angles. Top-view photographs are taken of the specimens during the loading process (Fig.~\ref{fig:S-ExpSetup}d), from which folding angles are measured using MATLAB.
The doubly-curved shape of the sheet upon longitudinal stretching is shown by the curved projected laser lines along the two midlines of the sheet (Figs \ref{fig:S-ExpSetup}e,f), where two orthogonal laser beams are projected from a laser ruler placed above and to the side of a loaded sheet, as shown by the schematics.

\section{Experiments on crumpled paper}
Experimental runs for three different crumpled sheets of paper are conducted using the same apparatus as for the plastic sheets. Sheets of printer paper ($80\mathrm{~gsm}$) of dimensions $W=8\mathrm{~cm}$, $L=16\mathrm{~cm}$ are first firmly crumpled by hand and are then loaded uniaxially with $\Wload= 1.6\mathrm{~cm}$. 

The angle $\alpha$ is measured directly as a function of the imposed longitudinal displacement, $\Delta u_y$ (see Eq.~\eqref{eqn:Deltauy} in the main text), by measuring the distances between two markers with the initial distance of about $14\mathrm{~cm}$. This is then converted to the planar strain~$\epsplan$,  as in Eq.~\eqref{eqn:epsplan} of the main text. (It is not possible to measure strains locally using the strain gauge because the creases introduced by crumpling mean that the strain gauge cannot be glued tightly to the paper.) Inclination angles are measured in the same way as for plastic sheets---from needles attached at the middle of the long edges (refer to Fig.~\ref{fig:S-ExpSetup}d). 

\begin{figure*}[h]
    \centering
    \includegraphics[width=0.5\textwidth]{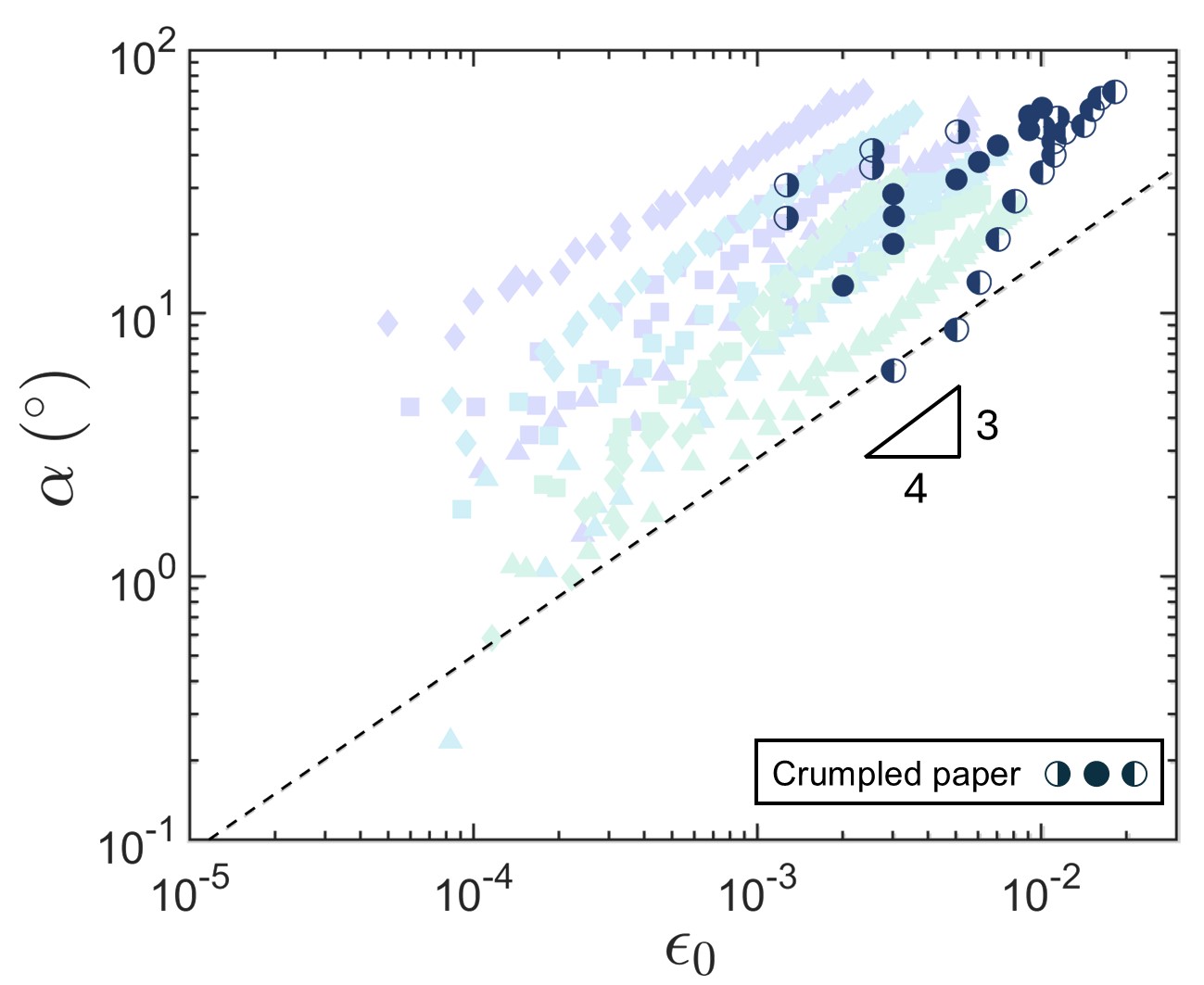}
    \caption{\textbf{Experimental data for crumpled paper sheets.} Results of tests on three crumpled-paper specimens (each with $L/W=2$, $\Wload/W=0.2$) as circles with different fills to distinguish each run. In this case, the measured strain is the planar strain $\epsplan$. Nevertheless, the observed folding angle generally displays a similar trend to that seen in experiments with plastic sheets (main text). The experimental data for plastic sheets is shown as the underlying transparent markers on the same plot for comparison, though we note that the strain measure for these data is actually $\epsloc$.}
    \label{fig:CrumpledSheets}
\end{figure*}

Results of these experiment on crumpled paper are shown in Fig.~\ref{fig:CrumpledSheets} for three repeats (each on a fresh sheet of paper). These data show the same qualitative trend as experiments on uniform elastic sheets presented in the main text; however, there is more scatter in the data, which we attribute to the manual, and hence less repeatable, way in which crumpling is introduced in the experiment. (Since it is unclear what the corresponding stretching and bending moduli are for each sheet, we present the data only in their raw form here; the raw results of experiments on plastic sheets are reproduced with increased transparency to facilitate comparison.) 

\section{Finite element method simulation} \label{sec:fem-method}
Numerical simulations are conducted using the commercial finite element analysis software ABAQUS/2023. The model is constructed in 3D space using 2D shell sections and S4R elements. Linear elastic constitutive relation is applied with Young’s modulus $E=3000\mathrm{~MPa}$ and with Poisson’s ratio $\nu\in\{0.01,0.3,0.49\}$. Two analysis steps are conducted in sequence: first, a small out-of-plane displacement, $\Delta u_z$, is applied to the midline at $x=0$ of a magnitude of $\sim$ 1\% of sheet thickness, creating initial perturbation to the planar sheet, while keeping both clamping edges fixed in all degrees of freedom; second, static analysis step is performed for the process of tensile-stretching induced folding of the elastic sheet. Displacement loading $\Delta u_y$ is applied to the bottom edge of the lower clamping region, while the top edge of the upper clamping region was fixed in all degrees of freedom.

\begin{figure*}[h]
    \centering
    \includegraphics[width=0.6\linewidth]{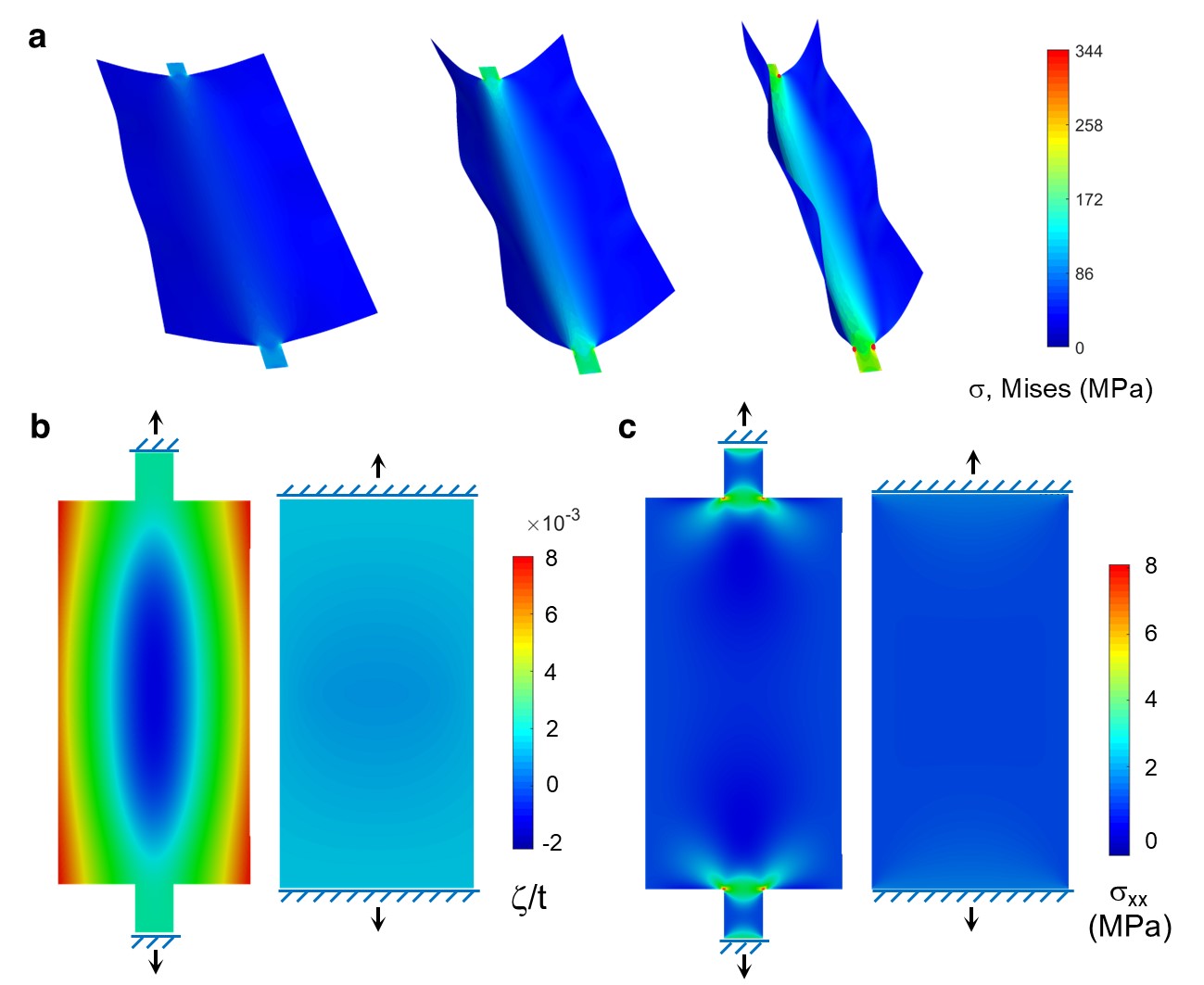}
    \caption{\textbf{FEM simulation results for the effect of loading width on stress distribution}.
    \textbf{a} Von-Mises stress distributions in sheets under localized tension, shown at $\epsbar=\mathrm{0.005,0.012,0.025}$. Sheet geometry is $\Wload/W=0.1,L/W=2,t/W=2.5\times10^{-3}$. 
    \textbf{b} Out-of-plane deflections, $\zeta$, normalized by thickness, $t$, for  sheets under localized loading (left), $\Wload/W=0.2$, and full-width loading (right), $\Wload/W=1$. 
    \textbf{c} Transverse stress, $\sigma_{xx}$, in the same simulations as \textbf{b} correspondingly. \textbf{b} and \textbf{c} have $L/W=2,t/W=5\times10^{-3}$, and are shown at the critical buckling strain for the localized loading case, i.e.~$\epsbar_{c}\approx 1.2 \times 10^{-3}$.}
    \label{fig:S-SimContourPlot}
\end{figure*}

\begin{figure*}[h]
    \centering
    \includegraphics[width=0.5\linewidth]{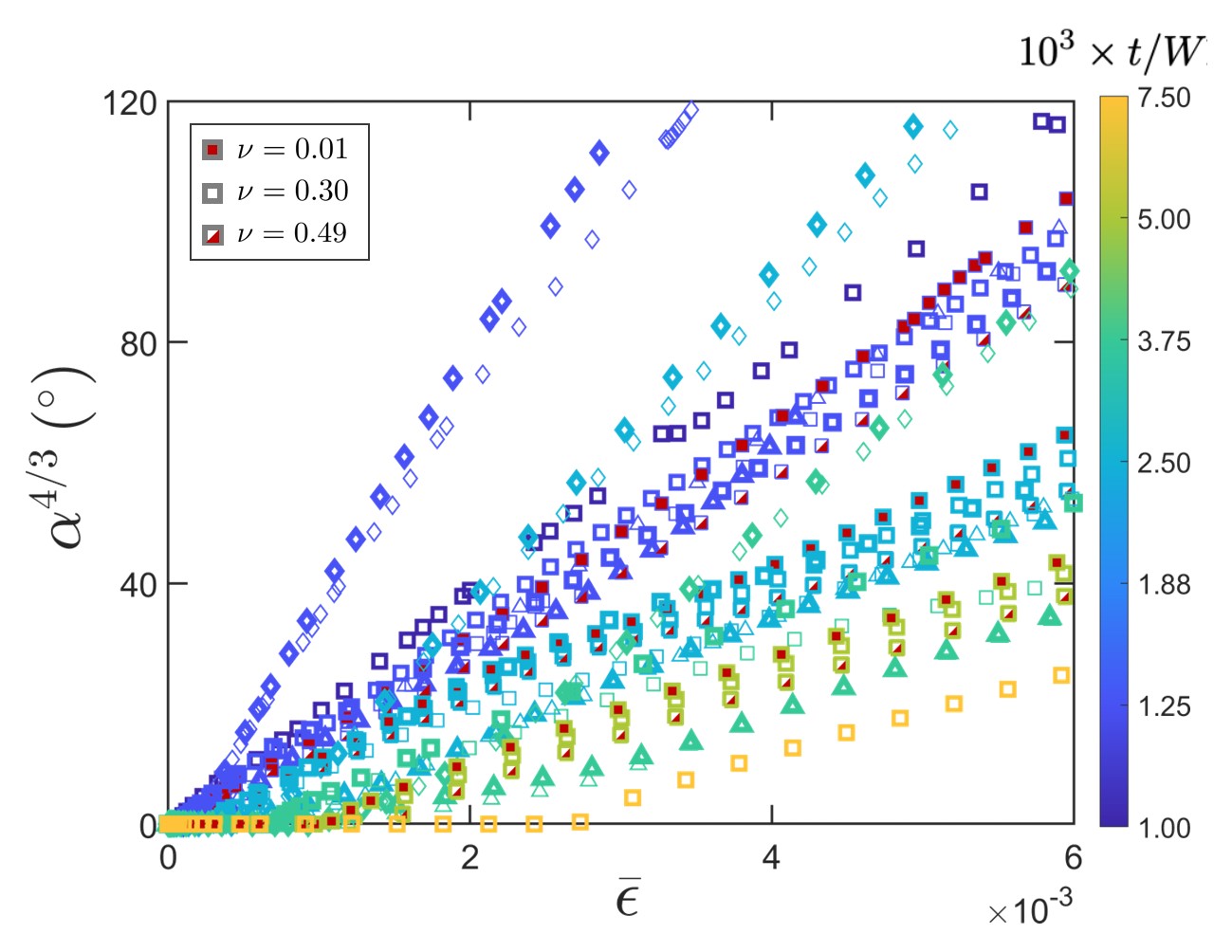}
    \caption{The scaled folding angle exhibits an intersection with the $x$-axis prior to the linear region, from where we determine the critical buckling strains $\epsbar_c$ for FEM simulations. Results from FEM simulations are presented with a variety of values of $L/W$ and $\Wload/W$ (given in the legends of Fig.~\ref{fig:FEMSimulation} in the main text), as well as $t/W$ (indicated by the color bar).}
    \label{fig:S-StrainAng}
\end{figure*}

\section{Comparison of different strains in experiments} \label{sec:Estimate-strain}

Our numerical simulations allow us to compare the various types of strain that are measured experimentally.

\subsection*{Local to mean strain}
\begin{figure*}[h]
    \centering
    \includegraphics[width=\linewidth]{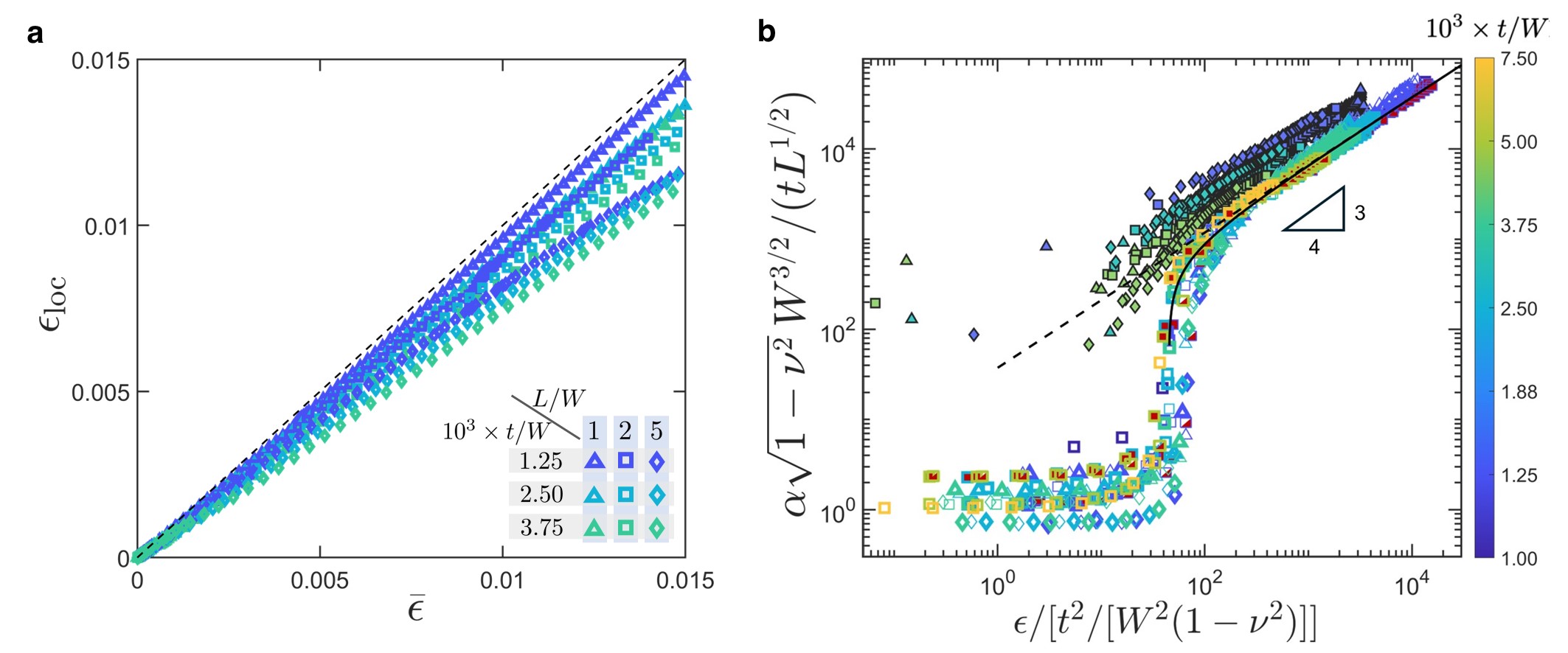}
    \caption{\textbf{a} Numerical simulation results of the local strain  at the centre of the sheet, $\epsloc$; to mimic the finite size of the strain gauge used experimentally, the strain is averaged over a longitudinal length  $l_{\mathrm{loc}}/W = 0.25$. Note that $\epsloc$ is smaller than the mean strain within the sheet, $\epsbar$, but remains an approximately linear function of $\epsbar$ throughout. In all cases shown  $\Wload/W=0.25$, corresponding to the parameter values used in the experiments. In the main text this relationship is used to convert the value of $\epsloc$ measured experimentally to an effective mean strain $\epsbar_{\mathrm{eff}}$. Without this conversion, the comparison between experimental and numerical results is slightly worse, as shown in panel \textbf{b}. Strain in the $x$-axis is $\epsbar$ for simulation data points, but is $\epsloc$ for experiment data points (as originally measured). The dashed line shows the expected power-law $\alpha\propto \epsbar^{3/4}$ for $\epsbar\gg\epsbar_c$;  the solid curve shows the scaling law modified to accommodate the critical behavior of the system, i.e.~Eq.~\eqref{eqn:InclinationAngle} of the main text.}
    \label{fig:S-LocGlob}
\end{figure*}

Firstly, our experiments measured the local strain, $\epsloc$, at the centre of the sheet. This can be converted to an effective mean strain, $\epsbar_{\mathrm{eff}}$, using numerical results, see Fig.~\ref{fig:S-LocGlob}a. However, even without this conversion, the comparison between experiments and numerical results, see Fig.~\ref{fig:S-LocGlob}b, is favourable.

\subsection*{Mean strain versus imposed longitudinal displacement}
\begin{figure*}[h]
    \centering
    \includegraphics[width=\linewidth]{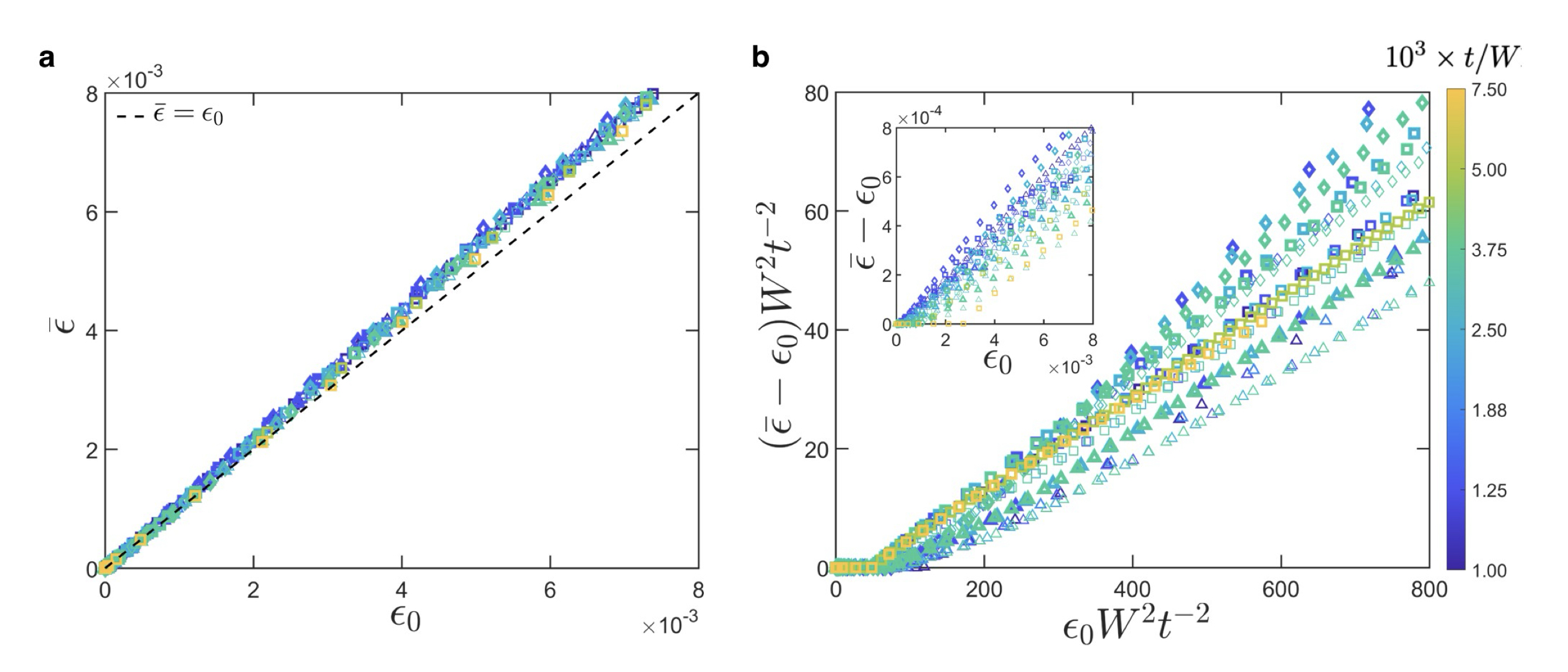}
    \caption{\textbf{a} Mean strain measured from the arclength, $\epsbar$, is larger than the directly imposed strain, $\epsplan\equiv$ $\Delta u_y/L$, in the post-buckling regime. \textbf{b} Up to the point of buckling $\epsbar=\epsplan$, but even after buckling there is a linear relationship $\epsbar-\epsplan \propto \epsplan-\epsbar_c$, as summarized in \eqref{eqn:StrainDispRelation}. Here the data are scaled so that all data sets pass through the buckling point, $(\epsbar_cW^2/t^2,0)$; the inset shows the data before scaling.}
    \label{fig:S-DeltayEpsbar}
\end{figure*}

Prior to buckling, the mean strain along the centre line $\epsbar=\epsplan=\Delta u_y/L$. However, after buckling, this relationship no longer holds. We therefore use numerical simulations to compute the relationship between the two: see Fig.~\ref{fig:S-DeltayEpsbar}a. In Fig.~\ref{fig:S-DeltayEpsbar}b we show that the discrepancy between the two grows linearly above the buckling threshold, so that:
    \begin{equation}
        \epsbar-\epsplan=\begin{cases}
        0,\quad \qquad \qquad \epsplan<\epsbar_c,\\
            k\left(\epsplan-\epsbar_c\right),\quad \epsplan>\epsbar_c.
        \end{cases}
        \label{eqn:StrainDispRelation}
    \end{equation}

Crucially, this means we can  use $\epsbar-\epsbar_c$ or $\epsplan-\epsbar_c$ interchangeably in scaling and experimental analyses. (Note, however, that the collapse of data is less satisfactory in the case $L=W$, for which the sheets are not particularly slender.)

\begin{figure}[ht]
    \centering
    \includegraphics[width=\linewidth]{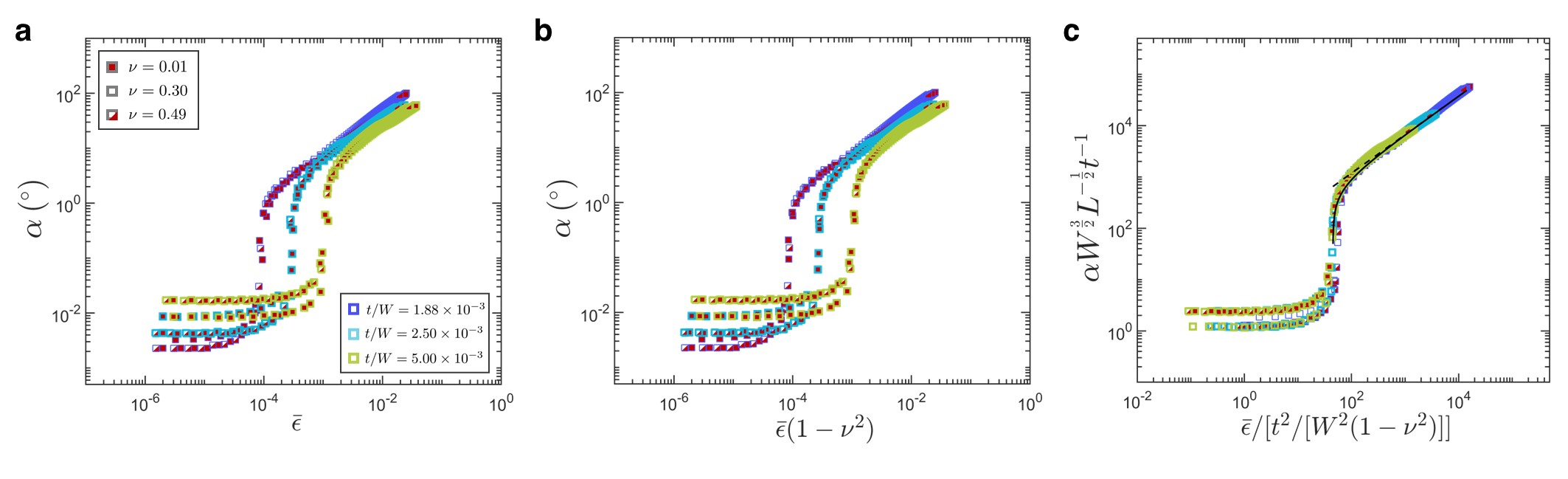}
    \caption{\textbf{Numerical simulation results for the inclination angle with different Poisson's ratios.} \textbf{a} Raw results showing the inclination angle as a function of strain with different thickness ratios (see legends). 
    \textbf{b} Rescaling the imposed strain by the Poisson's ratio-dependence in the critical strain provides some collapse of data with different $\nu$. 
    \textbf{c} Better collapse of the data is obtained by rescaling the $y$-axis as expected based on our theory. The dashed line shows the expected power-law $\alpha\propto \epsbar^{3/4}$ for $\epsbar\gg\epsbar_c$; the solid curve shows the scaling law modified to accommodate the critical behavior of the system, i.e.~Eq.~\eqref{eqn:InclinationAngle} of the main text.}
    \label{fig:poisson}
\end{figure}

\section{Details of Mathematical Modelling}\label{sec:Math-Model}
\begin{figure}[t!]
    \centering
    \begin{tikzpicture}
    \draw[draw=black] (0,0) rectangle ++(2,4);

    \draw [help lines,->] (1, 2) -- (1, 4.5);
    \draw [help lines,->] (1, 2) -- (2.5, 2);

    \draw [help lines,<->] (-0.25, 0) -- (-0.25, 4);
    \draw [help lines,<->] (0.75, 0.25) -- (1.25, 0.25);
    \draw [help lines,<->] (0, -0.25) -- (2, -0.25);

    \draw [draw=black,-] (1.25, 0) -- (1.2, -0.05);
    \draw [draw=black,-] (1.2, 0) -- (1.15, -0.05);
    \draw [draw=black,-] (1.15, 0) -- (1.1, -0.05);
    \draw [draw=black,-] (1.1, 0) -- (1.05, -0.05);
    \draw [draw=black,-] (1.05, 0) -- (1.0, -0.05);
    \draw [draw=black,-] (1.0, 0) -- (0.95, -0.05);
    \draw [draw=black,-] (0.95, 0) -- (0.9, -0.05);
    \draw [draw=black,-] (0.9, 0) -- (0.85, -0.05);
    \draw [draw=black,-] (0.85, 0) -- (0.8, -0.05);
    \draw [draw=black,-] (0.8, 0) -- (0.75, -0.05);
    \draw [draw=black,-] (0.75, 0) -- (0.7, -0.05);
    \draw [draw=black,->] (1, -0.1) -- (1, -0.2);
    \draw [draw=black,->] (1.2, -0.1) -- (1.2, -0.2);
    \draw [draw=black,->] (0.8, -0.1) -- (0.8, -0.2);

    \draw [draw=black,-] (1.25, 4.05) -- (1.2, 4);
    \draw [draw=black,-] (1.2, 4.05) -- (1.15, 4);
    \draw [draw=black,-] (1.15, 4.05) -- (1.1, 4);
    \draw [draw=black,-] (1.1, 4.05) -- (1.05, 4);
    \draw [draw=black,-] (1.05, 4.05) -- (1.0, 4);
    \draw [draw=black,-] (1.0, 4.05) -- (0.95, 4);
    \draw [draw=black,-] (0.95, 4.05) -- (0.9, 4);
    \draw [draw=black,-] (0.9, 4.05) -- (0.85, 4);
    \draw [draw=black,-] (0.85, 4.05) -- (0.8, 4);
    \draw [draw=black,-] (0.8, 4.05) -- (0.75, 4);
    \draw [draw=black,-] (0.75, 4.05) -- (0.7, 4);
    \draw [draw=black,->] (1, 4.1) -- (1, 4.2);
    \draw [draw=black,->] (1.2, 4.1) -- (1.2, 4.2);
    \draw [draw=black,->] (0.8, 4.1) -- (0.8, 4.2);

    \node at (2.75,2){$x$};
    \node at (1,4.75){$y$};
    \node at (-0.5,2){$L$};
    \node at (1,-0.5){$W$};
    \node at (1,0.5){$\Wload$};
    \node at (-1,4.5) {\textbf{a}};

    \end{tikzpicture}
    \begin{tikzpicture}
    \draw[draw=black] (1,0) rectangle ++(1,4);

    \draw [help lines,->] (1, 2) -- (1, 4.5);
    \draw [help lines,->] (1, 2) -- (2.5, 2);

    \draw [help lines,<->] (0.75, 0) -- (0.75, 4);
    \draw [help lines,<->] (1, -0.25) -- (2, -0.25);

    \draw [draw=black,->] (1, 0) -- (1, -0.2);
    \draw [draw=black,->] (1, 4) -- (1, 4.2);

    \node at (2.75,2){$x$};
    \node at (1,4.75){$y$};
    \node at (0.5,2){$L$};
    \node at (1.5,-0.5){$W/2$};
    \node at (0.75,-0.2){$T_{0}$};
    \node at (1.25,4.2){$T_{0}$};
    \node at (0,4.5) {\textbf{b}};

    \node at (4,4.5) {\textbf{c}};
    \end{tikzpicture}
    \includegraphics[width=0.4\linewidth]{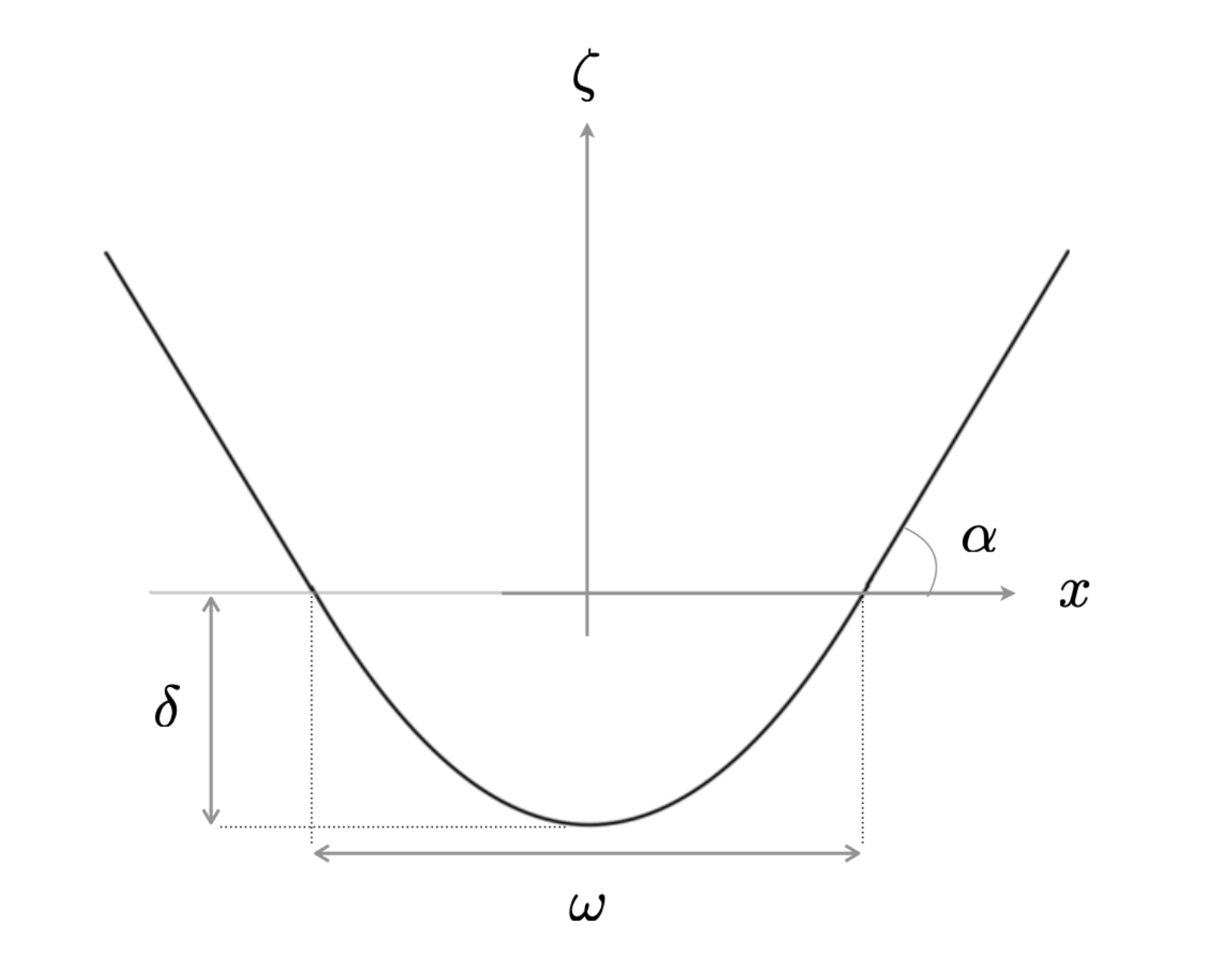}
    \caption{\textbf{Schematic diagrams.} \textbf{a} A rectangular sheet, of width $W$ and length $L$, clamped a width $\Wload<W$ along its width. The sheet is longitudinally loaded at the clamps. \textbf{b} A rectangular sheet, of width $W/2$ and length $L$, with localized load $T_{0}$ applied longitudinally at the corners $(0,\pm L/2)$. The sheet is transversely clamped at $(0,y)$. \textbf{c} A cross-sectional profile (at $y=0$) of the sheet stretched along the $y$-axis by a localized load beyond buckling. The origin in the $\zeta$-axis corresponds to the reference plane of the planar solution.}
    \label{fig:schematic}
\end{figure}

We consider an unstructured thin elastic sheet of thickness $t$, width $W$, and length $L$ ($t\ll W < L$) made of a material with Young's modulus $E$ and Poisson's ratio $\nu$.
The sheet is clamped in a central segment along the short edges, of width $\Wload$, and subjected to an in-plane longitudinal stretching strain $\epsilon_{yy}$ — we choose the long edges to be parallel to the $y$-axis. Outside the clamps (\emph{i.e.} for $|x|>\Wload/2$), the short edges are free, as are the long edges (see Fig.~\ref{fig:schematic}a).
We adopt the F\"oppl-von-K\'arm\'an (FvK) plate theory to model our sheet. We start by introducing the physical quantities that we will use. According to the FvK framework, we use the in-plane displacement-strain relationship:
\begin{align}
&\epsilon_{xx}=u_{x,x}+\frac{1}{2}(\zeta_{,x})^2;~\epsilon_{yy}=u_{y,y}+\frac{1}{2}(\zeta_{,y})^2,\nonumber\\
&\epsilon_{xy}=\frac{1}{2}\left(u_{y,x}+ u_{x,y}+\zeta_{,x}\zeta_{,y}\right),
\label{eq-strains}
\end{align}
where $u_x$, $u_y$ are in-plane components of the displacement field, whose out-of-plane component is $\zeta$ (we use the comma notation in the subindex to denote the derivatives).
We also invoke the two-dimensional Hookean constitutive relation, which links the in-plane strain to the in-plane stress ($\sigma_{\alpha\beta}$, which denotes the stress averaged across the sheet thickness and where indices $\alpha,\beta$ can be either $x$ or $y$):
\begin{align}
\sigma_{xx}=\frac{1}{1-\nu^2}Y\left(\epsilon_{xx}+\nu \,\epsilon_{yy}\right),\nonumber\\
\sigma_{yy}=\frac{1}{1-\nu^2}Y\left(\epsilon_{yy}+\nu\, \epsilon_{xx}\right),\nonumber\\
\sigma_{xy}=\frac{1}{1+\nu}Y\,\epsilon_{xy},
\label{eq-Hooke}
\end{align} where $Y=E\,t$ is the stretching modulus. Note that here the $2d$-stresses have dimensions of a tension (force $\times$ $\text{length}^{-1}$) --- in FvK, the in-plane stresses averaged across the thickness are $t\times\sigma_{\alpha\beta}(z=0)$.

We consider the limit of small $\Wload$, $\Wload/W\ll1$. Exploiting the symmetry along the $x$-axis, that we align with the short edge, we consider one half of this sheet, \emph{i.e.} a flat sheet of width $W/2$ and length $L$ (see Fig.~\ref{fig:schematic}b). We examine next the instability of this sheet through a near-threshold analysis~\cite{XinDavidovitch21} of the planar state (i.e. $\zeta=0$). For the beyond threshold analysis, whose main parameters are represented in Fig.~\ref{fig:schematic}c, see the main text.

\subsection*{The planar state}
The FvK equations that set the mechanical equilibrium in the planar state reduce to a single bi-harmonic equation:
\begin{align}
\nabla^4\chi(x,y)=0,
\label{eq-biharmonic}
\end{align}
after we eliminate the in-plane components of the stress tensor in favour of a single scalar function, the Airy potential $\chi(x,y)$, defined such that
\begin{align}
\sigma_{xx}=\chi_{,yy};\, \sigma_{yy}=\chi_{,xx};\, \sigma_{xy}=-\chi_{,xy}.
\label{eq-Airy-def}
\end{align}

We consider only small strains, and hence we assume the boundary conditions (BCs) to apply at the edges $x=0,W/2$ and $y=\pm L/2$. 
For convenience, we  rescale $x,y$ by $L/2$ so that the edges are at $y=\pm 1$. 
The set of BCs for the stress and the displacement fields is then:
\begin{subequations}
\begin{align}
&\text{at $y=\pm 1$, $x\neq0$:}&\, \sigma_{yy}=\sigma_{xy}=0,\\
&\text{at $x=W/2$:}&\, \sigma_{xx}=\sigma_{xy}=0,\label{eq-BCs-stress-displ-atyW}\\
&\text{at $x=0$:}&\, u_x=0,\,u_{y,y}=\epsplan,\label{eq-BCs-stress-displ-aty0}
\end{align}
\label{eq-BCs-stress-displ}
\end{subequations}
which describes a sheet whose edge at $x=0$ is both transversely clamped ($u_x(x=0)=0$) and longitudinally strained ($\epsilon_{yy}(x=0)=\Delta u_y/L\equiv\epsplan>0$, with $\Delta u_y\equiv u_y(x=0,y=L/2)-u_y(x=0,y=-L/2)$). The edges at $x=W/2$ and $y=\pm 1$ are free.

We scale the stresses by the characteristic tension $T_{0}=Y\epsplan$ (with $Y=Et$), and so let $\sigma_{\alpha\beta}/T_{0}$ be the dimensionless stress tensor. Converting the set of BCs~\eqref{eq-BCs-stress-displ} into BCs for the Airy potential, via \eqref{eq-Airy-def}, we find:
 \begin{subequations}
\begin{align}
\text{at $y=\pm 1$, $x\neq0$:}&\quad \chi=0\quad\text{and}\quad\chi_{,y}=0,\label{eq-BC-Airy-at-x}\\
\text{at $x=W/2$:}&\quad \chi=0\quad\text{and}\quad\chi_{,x}=0,\label{eq-BC-Airy-at-yW}\\
\text{at $x=0$:}&\quad \chi_{,xx}-\nu\chi_{,yy}=1\quad\text{and} \quad \chi_{,xxx}+(2+\nu)\chi_{,xyy}=0.\label{eq-BC-Airy-at-y0}
\end{align}
\label{eq-BCs-Airy}
 \end{subequations} (Note that we do not introduce any different notation for the dimensionless stresses $\sigma_{\alpha\beta}$, nor for the corresponding Airy potential $\chi$, but work in dimensionless terms henceforth.)
Both equations in~\eqref{eq-BC-Airy-at-x} and~\eqref{eq-BC-Airy-at-yW} follow from choosing the natural gauge of vanishing constants of integration when we integrate respectively the first equation in~\eqref{eq-Airy-def} at $y=\pm 1$ and the second at $x=W/2$.
This gauge is consistent with the BCs for the off-diagonal stress components at the free edges $y=\pm 1$ and $x=W/2$.

By virtue of the small strains approximation, viz.~\eqref{eq-strains}, the second BC in~\eqref{eq-BCs-stress-displ-aty0} implies that $\epsilon_{yy}(x=0)=\epsplan$.
Invoking the inverse of the Hookean constitutive relations~\eqref{eq-Hooke}, and using the definition of the Airy potential~\eqref{eq-Airy-def}, the second BC in~\eqref{eq-BCs-stress-displ-aty0} converts into the first equation in~\eqref{eq-BC-Airy-at-y0}. Finally, to obtain the last BC in~\eqref{eq-BC-Airy-at-y0} we use the off-diagonal strain, $\epsilon_{yx}=\sigma_{yx}\,(1+\nu)/Y$ by inverting the last relation in~\eqref{eq-Hooke}, which evaluates to $\epsilon_{yx}=u_{y,x}/2$ at the edge $x=0$ after we apply the BCs~\eqref{eq-BCs-stress-displ-aty0} into~\eqref{eq-strains}. Thus, we have $u_{y,x}/2=\sigma_{yx}\,(1+\nu)/Y$. Writing this equation in terms of the Airy potential~\eqref{eq-Airy-def}, and taking its $y$ derivative yields the equation $u_{y,xy}=-\chi_{,yyx}\,2(1-\nu)/Y$ for the $x,y$ derivative of the longitudinal displacement. The latter can be obtained from the $x$ derivative of $\epsilon_{yy}$: $u_{y,xy}=\epsilon_{yy,x}=(\chi_{,xxx}-\nu\,\chi_{,yyx})/Y$.  The second equation in~\eqref{eq-BC-Airy-at-y0} then follows from eliminating $u_{y,xy}$ from the last two equations.

We solve the bi-harmonic equation~\eqref{eq-biharmonic} for the Airy potential with the set of BCs~\eqref{eq-BCs-Airy} following the calculation of Benthem~\cite{Benthem63} (further details on the method are given in the next subsection). In short, the general solution may be written
\begin{align}
\chi(x,y)=\text{Re}\Big[\sum_k\Big\{{\displaystyle C}_k\,e^{p_k(x-W/2)}+\hat{\displaystyle C}_k\,e^{-p_k x}\Big\}\Big\{\cos \left(p_k y\right)-(\cot p_k)y\,\sin\left(p_k y\right)\Big\}\Big]\,,\label{eq-Airy-final}
\end{align}
where ${\displaystyle C}_k$, $\hat{\displaystyle C}_k$ are constants given in~\eqref{eq-coeffs-Airy} in Sec.~\ref{sec:Benthem-method}, and $p_k$ are the (complex) roots of the trigonometric equation
\begin{align}
\sin 2p_k+2p_k=0,
\label{eq-poles}
\end{align}
ordered by increasing $|p_k|$.
Differentiating the Airy potential, we obtain the transverse and longitudinal components of the stress tensor~\eqref{eq-Airy-def}:
\begin{subequations}
\begin{align}
\sigma_{xx}=\chi_{,yy}=&\text{Re}\Big[\sum_k\Big\{{\displaystyle C}_k\,e^{p_k(x-W/2)}+\hat{\displaystyle C}_k\,e^{-p_k x}\Big\}\Big\{-\left(p_k^2+2p_k\cot p_k\right)\cos \left(p_k y\right)+\cot p_k\,y\,\sin\left(p_k y\right)\Big\}\Big]\,,\label{eq-transv-stress}\\
\sigma_{yy}=\chi_{,xx}=&\text{Re}\Big[\sum_k\left(p_k\right)^2\Big\{{\displaystyle C}_k\,e^{p_k(x-W/2)}+\hat{\displaystyle C}_k\,e^{-p_k x}\Big\}\Big\{\cos \left(p_k y\right)-(\cot p_k)y\,\sin\left(p_k y\right)\Big\}\Big]\,.\label{eq-long-stress}
\end{align}
\label{eq-stress}
\end{subequations}

\begin{figure}[t!]
    \centering
    \includegraphics[width=\linewidth]{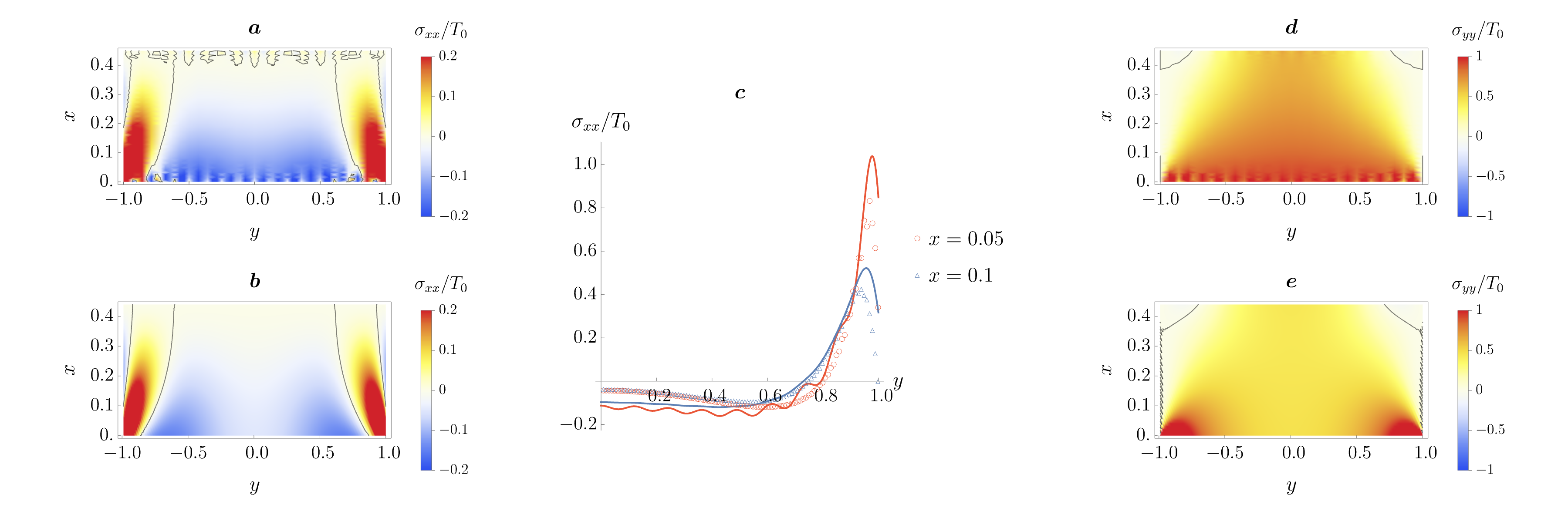}
    \caption{\textbf{Stress in the planar state.} \textbf{a} Analytic solution for the transverse stress~\eqref{eq-transv-stress} within the sheet subjected to a longitudinal localized load at the corners as per Fig.~\ref{fig:schematic}b; $W/L=0.45$ and we take $n=16$ terms in the series. \textbf{b} Numerical results for the transverse stress obtained by finite-element simulations  of a sheet ($L/W=2,~t/W=0.0025,~\Wload/W=0.1,~E=3000~\text{MPa},~\nu=0.3$) subjected to a longitudinal strain $\epsplan=0.025$. The numerical results are rescaled according to the scaling in the mathematical model, and they are only shown within $x\geq \Wload/2$. The same colorbar as in \textbf{a} applies. \textbf{c} Cross-sectional profiles of the transverse stress at different values of $x$; symbols for the numerics, lines for the analytic expression. \textbf{d} Longitudinal component of the stress as predicted by~\eqref{eq-long-stress}. \textbf{e} Numerical results for the longitudinal stress with the same colorbar as in \textbf{d}.
    }
    \label{fig:PlanarStressState}
\end{figure}

We benchmark the expressions for the stress in the planar state~\eqref{eq-stress} with the finite-element simulations  of a sheet ($L/W=2,~t/W=2.5\times10^{-3},~\Wload/W=0.1,~E=3000~\text{MPa},~\nu=0.3$) subjected to longitudinal strain $\epsplan=0.025$. Considering the scaling assumed previously (\emph{i.e.} lengths are rescaled by $L/2$ and $L=2$, and stresses are normalized by $T_0=Y\,\epsplan$), we plot in Fig.~\ref{fig:PlanarStressState} the planar stress state obtained from~\eqref{eq-stress} and the numerical simulations (for which we consider only the region $x\geq\Wload/2$). We notice a good agreement between the predictions of the analytic model and the corresponding numerical simulations: compare Fig.~\ref{fig:PlanarStressState}a with Fig.~\ref{fig:PlanarStressState}b (transverse stress) and Fig.~\ref{fig:PlanarStressState}d with Fig.~\ref{fig:PlanarStressState}e (longitudinal stress). The transverse stress (Fig.~\ref{fig:PlanarStressState}a and b) is tensile at the corners where the characteristic tension is $T_0$, and compressive as we move away from the corners towards the centre --- we have included the contour line at $\sigma_{xx}/T_{0}=0$ to evince the coexistence of compressive and tensile areas in the sheet. Close to the $y$-axis, the cross-sectional profiles of this stress component (Fig.~\ref{fig:PlanarStressState}c) help us visualize the tension-to-compression transition as we move away from the corners, and the stress singularity at their vicinity (see the next section for more details on the stress singularities at the corners of a sheet). Away from the $y$-axis, the transverse stress within the sheet fades rapidly. Conversely, the longitudinal stress (Fig.~\ref{fig:PlanarStressState}d and e) is tensile everywhere in the sheet, and hence is given less attention in our discussion.

We further examine the transverse stress for different values of the aspect ratio $W/L$. We already noticed that the transverse stresses are weak away from the corners of the sheet that are loaded. Thus, we focus on the stress state close to the loaded corners.
We plot in Fig.~\ref{fig:PlanarStressTransverse} the expression~\eqref{eq-transv-stress} shifted in $y$, $y\to y+1$, and we scale both axes by $W/2$. We observe a universal stress state for $W/L\lesssim 1/4$: the transverse stress is tensile at the vicinity of the corner, and compressive within an area that borders the narrow tensile part. The width of this compressing sector scales with $\sim W$ --- we have included contour lines every $\Delta(\sigma_{xx}\/T_0)=-0.05$ in the compressive part so as to evince the behavior of these  stresses. This scaling is paramount in our analysis in the main text. For $W/L\gtrsim 1/4$, the compressive sectors corresponding to both corners overlap, and hence the sheet has a single compressed area spanning the centre of the sheet, while the two tensile sectors prevail at the corners.

\begin{figure}[t!]
    \centering
    \includegraphics[width=\linewidth]{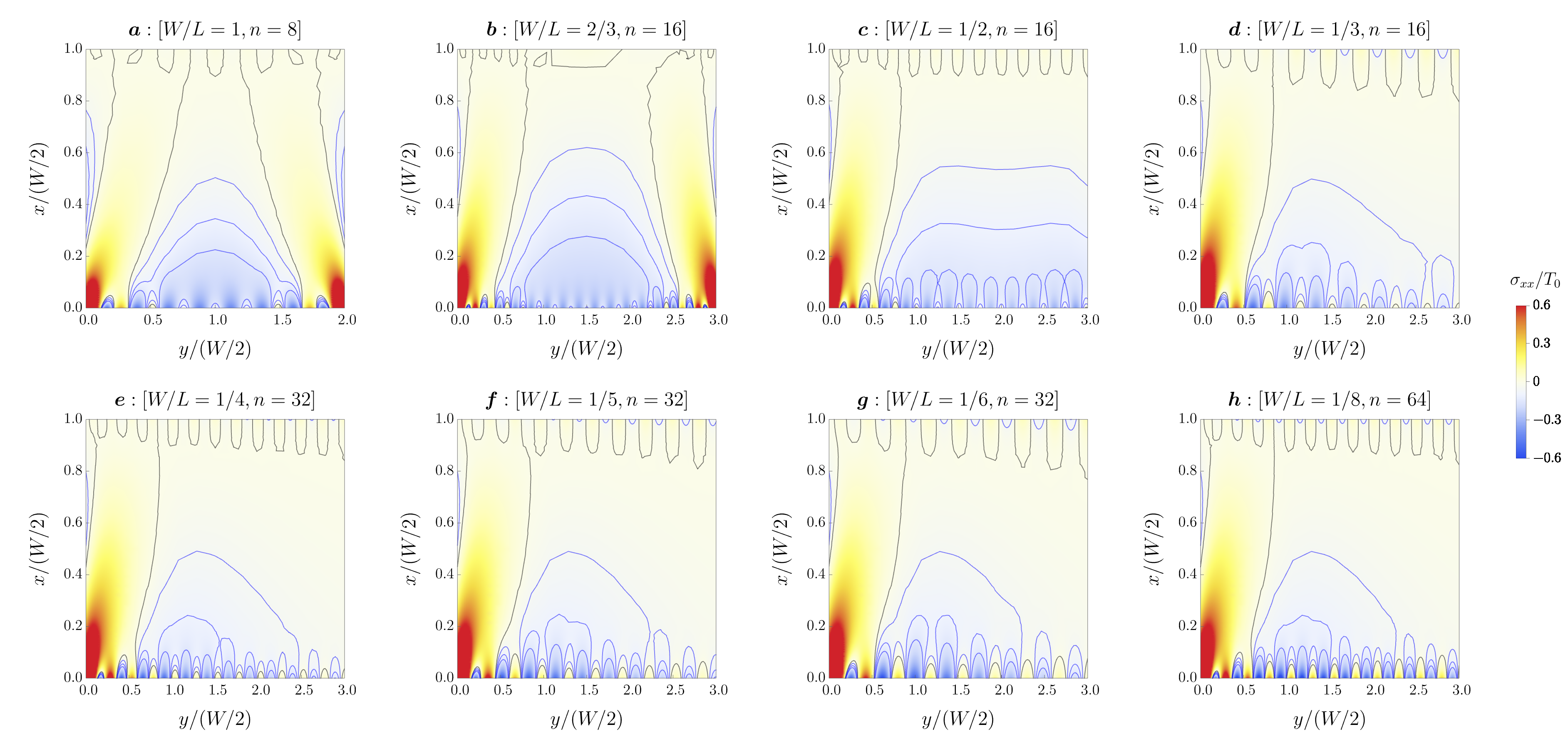}
    \caption{\textbf{Transverse stress in the planar state for different aspect ratios.} \textbf{a-h} Density plots of the stress by the expression~\eqref{eq-transv-stress}. The longitudinal coordinate, $y$, is shifted $y\to y+1$; and both axes are scaled by $W/2$. The same colorbar (only shown in \textbf{a}) applies to all panels. The number of terms, $n$, taken in the series~\eqref{eq-transv-stress} is labeled in each graph. Contour lines are displayed at $\sigma_{xx}/T_{0}=(0,-0.05,-0.1,-0.15)$.
    }
    \label{fig:PlanarStressTransverse}
\end{figure}

\subsection*{Benthem's Laplace transform method} \label{sec:Benthem-method}
We define the function $f(y,p)$ as the Laplace transformation of the Airy potential,
\begin{align}
f(y,p)=\int_0^{W/2}\chi(x,y)\, e^{-px}\,\mathrm{d}x,
\label{eq-Laplace}
\end{align}
and the subsequent inverse formula
\begin{align}
\chi(x,y)=\frac{1}{2\pi\mathrm{i}}\int_{c-\mathrm{i}\infty}^{c+\mathrm{i}\infty}f(y,p)\,e^{px}\,\mathrm{d}p\quad(c\in\mathbb R, c>0).
\label{eq-Laplace-inv}
\end{align}

The Laplace transformation of the bi-harmonic equation~\eqref{eq-biharmonic} is
\begin{align}
p^4\,f(y,p)+2\,p^2\,f_{,yy}(y,p)+f_{,yyyy}(y,p)=&
p^3\,\chi|_{x=0}+\,p^2\chi_{,x}|_{x=0}-p\left[-2\chi_{,yy}|_{x=0}+e^{-px}\chi_{,xx}|^{  x=W/2}_{x=0}\right]\nonumber\\
&+2\chi_{,yyx}|_{x=0}-e^{-px}\chi_{,xxx}|^{x=W/2}_{x=0},
\label{eq-biharmonic-Laplace}
\end{align} 
where we have already substituted the BCs at $x=W/2$~\eqref{eq-BC-Airy-at-yW}. Notice that we cannot infer all the boundary terms in the \emph{rhs} of the equation solely from the Airy potential BCs~\eqref{eq-BCs-Airy}, and so we express the unknown functions with the aid of Fourier series. However, these terms should account for the stress singularities in the corners of a sheet when one edge is clamped and the other is free~\cite{Williams52} (see~\cite{Benthem63} and~\cite{XinDavidovitch21} for a complete calculation of the stress singularities in the vicinity of a sheet corner with clamped-free edges). For $\nu\sim 0.32$, $\sigma_{\theta\theta}$ and $\sigma_{\theta r}$ are given by~\cite{Benthem63}:
\begin{align}
\sigma_{\theta\theta}\left(\theta=\frac{\pi}{2}\right)= c_1 r^{-1/4},\qquad \sigma_{r\theta}\left(\theta=\frac{\pi}{2}\right)=-c_2 r^{-1/4},
\end{align}
where $c_1=\sqrt{8+3\sqrt{7}}c_2$, and $c_2$ is a constant to be determined. Thus, in the corners of our sheet ($x=0,y=\pm 1$) we assume
\begin{subequations}
\begin{align}
\sigma_{xx}(x=0,y\to \pm 1)=c_1 (1\mp y)^{-1/4},\\
\sigma_{xy}(x=0,y\to \pm 1)=\mp c_2 (1\mp y)^{-1/4},
\end{align}
\label{eq-stress-singularities}
\end{subequations}
\noindent where the choice for the signs follows from the following discussion. When stretching longitudinally the transversely clamped edge ($x=0$), we exert a tension onto the sheet, and hence $\sigma_{xy}(x=0,y\to \pm 1)$ should be pointing outwards at both corners: $\sigma_{xy}(x=0,y\to \pm 1)\to \mp\infty$ (notice that the surface's unit vector carries an extra minus sign in the definition of the component of the stress tensor as it points downwards in the $x$-direction). Along the transverse ($x$) direction, the transverse clamp at $x=0$ prevents Poisson's effect --- by which a longitudinal stretching induces transverse contraction --- and hence it exerts a tensile transverse stress, thus $\sigma_{xx}(x=0,y\to \pm 1)\to +\infty$.

Hence, invoking the stress singularities at the corners~\eqref{eq-stress-singularities}, we write
\begin{subequations}
\begin{align}
\chi_{,yy}|_{x=0}=\sigma_{xx}(0,y)=c_1 (1-y)^{-1/4}+c_1 (1+y)^{-1/4}-c_1 2^{-1/4}+\sum_m a_m \cos\left(\frac{m\pi}{2}y\right),\\
\chi_{,xy}|_{x=0}=-\sigma_{xy}(0,y)=c_2 (1-y)^{-1/4}-c_2 (1+y)^{-1/4}+c_2 2^{-1/4}y+\sum_n b_n \sin\left(\frac{n\pi}{2}y\right),
\end{align}
\end{subequations} where $a_m$ and $b_n$ are yet to be determined constants, and $m=2k-1$, $n=2k$ for $k\in{\mathbb Z}^+$. We obtain from these the remaining unknown functions in~\eqref{eq-biharmonic-Laplace}:
\begin{subequations}
\begin{align}
\chi_{,xx}|_{x=0}=&1 +\nu\, \chi_{,yy}|_{x=0},~\text{from}\,\eqref{eq-BC-Airy-at-y0},\\
\chi_{,xyy}|_{x=0}=&-\sigma_{xy,y}(0,y)=\frac{c_2}{4} (1-y)^{-5/4}+\frac{c_2}{4} (1+y)^{-5/4}+c_2 2^{-1/4}+\frac{\partial}{\partial y}\sum_n b_n \sin\left(\frac{n\pi}{2}y\right),\\
\chi_{,xxx}|_{x=0}=&-(2+\nu)\chi_{,xyy},~\text{from}\,\eqref{eq-BC-Airy-at-y0},\\
\chi_{,x}|_{x=0}=&\int\mathrm{d}y\,\chi_{,xy}|_{x=0}+b_0=b_0-\frac{4}{3}c_2 (1-y)^{3/4}-\frac{4}{3}c_2(1+y)^{3/4}+c_2 2^{-5/4}y^2+\sum_n b_n\left(\frac{-2}{n\pi}\right) \cos\left(\frac{n\pi}{2}y\right),\\
\chi|_{x=0}=&\int\mathrm{d}y\,\left(\int\mathrm{d}\hat y\,\chi_{,\hat y\hat y}|_{x=0}+c_4\right)+c_5\nonumber\\
=&c_5+c_4 y+\frac{16}{21}c_1(1-y)^{7/4}+\frac{16}{21}c_1(1+y)^{7/4}-c_1 2^{-5/4}y^2-\sum_m a_m \left(\frac{2}{m\pi}\right)^2\cos\left(\frac{m\pi}{2}y\right),
\end{align}
\end{subequations} where $c_4=0$, $c_5=c_1\left(-\frac{16}{21}2^{7/4}+2^{-5/4}\right)$ are computed using the BCs~\eqref{eq-BC-Airy-at-x}. Finally, we use the \emph{cosine} Fourier series for $\chi_{,xx}|_{x=W/2}$ and $\chi_{,xxx}|_{x=W/2}$,
\begin{subequations}
    \begin{align}
    \chi_{,xx}|_{x=W/2}=&\sum_m\alpha_m\cos\left(\frac{m\pi}{2}y\right),\\
    \chi_{,xxx}|_{x=W/2}=&\sum_m\beta_m\cos\left(\frac{m\pi}{2}y\right),
    \end{align}
\end{subequations}$m=2k-1$ for $k\in{\mathbb Z}^+$, where we exploited the symmetry in $y$, and we write the transformed bi-harmonic equation:
\begin{align}
p^4\,f(y,p)+2\,p^2\,f_{,yy}(y,p)+f_{,yyyy}(y,p)=&p^3c_1\left[\frac{16}{21}(1-y)^{7/4}+\frac{16}{21}(1+y)^{7/4}-2^{-5/4}y^2+2^{-5/4}-\frac{16}{21}2^{7/4}\right]\nonumber\\
+&p^2\left\{c_2\left[-\frac{4}{3}(1-y)^{3/4}-\frac{4}{3}(1+y)^{3/4}+2^{-5/4}y^2\right]+b_0\right\}\nonumber\\
+&p\left\{c_1(2+\nu)\left[(1-y)^{-1/4}+(1+y)^{-1/4}-2^{-1/4}\right]+1\right\}\nonumber\\
-&\nu c_2\left[\frac{1}{4}(1-y)^{-5/4}+\frac{1}{4}(1+y)^{-5/4}+2^{-1/4}\right]\nonumber\\
-&p^3\sum_{m}a_m\left(\frac{2}{m\pi}\right)^2\cos\left(\frac{m\pi}{2}y\right)-p^2\sum_{n}b_n\frac{2}{n\pi}\cos\left(\frac{n\pi}{2}y\right)\nonumber\\
+&p\left[(2+\nu)\sum_{m}a_m\cos\left(\frac{m\pi}{2}y\right)-e^{-p\,W/2}\sum_{m}\alpha_m\cos\left(\frac{m\pi}{2}y\right)\right]\nonumber\\
-&\nu\frac{\partial}{\partial y}\sum_{n}b_n\sin\left(\frac{n\pi}{2}y\right)-e^{-p\,W/2}\sum_{m}\beta_m\cos\left(\frac{m\pi}{2}y\right),
\end{align} $m=2k-1$, $n=2k$ for $k\in{\mathbb Z}^+$.

We solve this differential equation for $f(y,p)$ and the corresponding BCs~\eqref{eq-BC-Airy-at-x}, which become
\begin{align}
f(\pm 1,p)=0;~f_{,y}(y,p)|_{y=\pm 1}=0,
\end{align}
and we write the solution $f(y,p)$ as
\begin{align}
f(y,p)=&f_p(y,p)-\frac{2\,f_p(1,p)}{\sin 2p +2p}\left[(\sin p+p\cos p)\cos py+(p\sin p)y\sin py\right]\nonumber\\
+&\frac{2\,f_{p,y}(1,p)}{\sin 2p+2p}\left[\sin p\cos py-(\cos p)y\sin py\right].
\label{eq-sol-fyp}
\end{align}
Here the particular integral
\begin{align}
f_p(y,p)=&p^3 c_1\frac{16}{21}\left\{R\left[\frac{7}{4};p,-y\right]+R\left[\frac{7}{4};p,y\right]\right\}-p^2 c_2\frac{4}{3}\left\{R\left[\frac{3}{4};p,-y\right]+R\left[\frac{3}{4};p,y\right]\right\}\nonumber\\
+&p c_1(2+\nu)\left\{R\left[\frac{-1}{4};p,-y\right]+R\left[\frac{-1}{4};p,y\right]\right\}-\frac{1}{4}\nu c_2\left\{R\left[\frac{-5}{4};p,-y\right]+R\left[\frac{-5}{4};p,y\right]\right\}\nonumber\\
+&\frac{2^{-5/4}}{p^4}(-p c_1+c_2)(y^2p^2-4)-\frac{2^{-1/4}}{p^4}\nu c_2+\frac{1}{p^3}[-2^{-1/4}c_1(2+\nu)+1]\nonumber\\
+&\frac{1}{p^2}b_0+\frac{1}{p}c_1\left(2^{-5/4}-\frac{16}{21}2^{7/4}\right)+\sum_m\frac{a_m\left[-p^3\left(\frac{2}{m\pi}\right)^2+p(2+\nu)\right]-e^{-p W/2}(p\,\alpha_m+\beta_m)}{\left[p^2-\left(\frac{m\pi}{2}\right)^2\right]^2}\cos\left(\frac{m\pi}{2}y\right)\nonumber\\
-&\sum_n b_n\frac{p^2\frac{2}{n\pi}+\nu\frac{n\pi}{2}}{\left[p^2-\left(\frac{n\pi}{2}\right)^2\right]^2}\cos\left(\frac{n\pi}{2}y\right),
\label{eq-particular-integral}
\end{align} $m=2k-1$, $n=2k$ for $k\in{\mathbb Z}^+$; and, to express this more compactly, we defined an `auxiliary' function $R[q;p,y]$:
\begin{align}
R[q;p,\pm y]\equiv \frac{1}{2 p^3}\int_0^y(1\pm\xi)^q\sin p(y-\xi)\, \mathrm{d}\xi-\frac{1}{2p^2}\int_0^y(1\pm\xi)^q (y-\xi)\cos p(y-\xi)\, \mathrm{d}\xi .
\end{align}

To obtain the Airy potential, $\chi(x,y)$, we compute the inverse Laplace transform~\eqref{eq-Laplace-inv} of the function $f(y,p)$~\eqref{eq-sol-fyp}. To calculate the line integral~\eqref{eq-Laplace-inv}, we define a closed contour $\mathcal{C}$ in the complex plane and then apply the Residue Theorem~\cite{MorseFeshbachI}. Note that, for arbitrary values of $(c_1, b_0, a_m, b_n, \alpha_m, \beta_m)$, the function $f(y,p)$ has poles at $p=p_k$, where $p_k$ are the (complex) roots of the trigonometric equation~\eqref{eq-poles} ($p_k\neq 0$) --- the double poles at $p=\pm n\pi/2$, $p=\pm m\pi/2$, and the pole at $p=0$ can be removed. Thus, in order that the integral~\eqref{eq-Laplace-inv} converges we need these poles at $p=p_k$ to be removed, \emph{i.e.} their residues to vanish
\begin{align}
\mathrm{Res}[f(y,p),p=p_k]=\frac{1}{2}[f_{p,y}(y,p_k)|_{y=1}-\sin^2 p_k f_p(1,p_k)]\frac{[\sin p_k\cos p_k y-(\cos p_k)y\sin p_k y]}{\cos^2 p_k}=0.
\label{eq-residues}
\end{align} This leads to an infinite system of algebraic equations in the unknowns $(c_1, b_0, a_m, b_n, \alpha_m, \beta_m)$. Following the calculation of Benthem~\cite{Benthem63}, we take only a finite number of unknowns --- \emph{i.e.} we truncate the Fourier series in the particular integral~\eqref{eq-particular-integral}; the  number of equations sufficient to calculate $(c_1, b_0, a_m, b_n, \alpha_m, \beta_m)$ is then obtained by using the necessary number of values of $p_k$ (ordered by increasing real part). Substituting the numerical values of $(c_1, b_0, a_m, b_n, \alpha_m, \beta_m)$ into the particular integral~\eqref{eq-particular-integral}, we obtain a function $f(y,p)$ that has no poles, and thus whose complex line integral~\eqref{eq-Laplace-inv} converges.

So far our discussion has focused on the function $f(y,p)$, yet we notice that it multiplies the exponential $e^{px}$ in the inverse Laplace formula~\eqref{eq-Laplace-inv}. We close the integration contour so that $f(y,p)e^{px}$ is an analytic function, continuous within $\mathcal{C}$. Hence, by writing $f(y,p)$ as
\begin{subequations}
\begin{align}
f(y,p)=e^{-p W/2}f_{+}(y,p)+f_-(y,p)&,\\
\text{with}\qquad f_+(y,p)=&\sum_m \frac{(p\alpha_m+\beta_m)}{\left[p^2-\left(\frac{m\pi}{2}\right)^2\right]^2}\Bigg[-\cos\left(\frac{m\pi}{2}y\right)\nonumber\\
+&m\pi\sin\left(\frac{m\pi}{2}\right)\frac{\left[\sin p\cos py-(\cos p)y\sin py\right]}{\sin 2p+2p}\Bigg],\\
\text{and} \qquad f_-(y,p)=&f(y,p)-e^{-p W/2}f_+(y,p),
\end{align}\label{eq-redef-f} \end{subequations} $m=2k-1$ for $k\in{\mathbb Z}^+$, and substituting this into the integral~\eqref{eq-Laplace-inv}, we find that the integration line must be closed in either of the contours $\mathcal{C}_{\pm}$ in Fig.~\ref{fig:contours} for the integral of each term. Note that the functions $f_{\pm}(y,p)$ in~\eqref{eq-redef-f} have poles at $p=p_k$.
Applying the Residue Theorem~\cite{MorseFeshbachI} (or Cauchy's Theorem in those cases for which the integrand is $f(y,p)e^{px}$), we find that
\begin{align}
\chi(x,y)=\begin{cases}
\frac{1}{2\pi\mathrm{i}}\oint\limits_{\mathcal{C}_+}f(y,p)e^{px}\,\mathrm{d}p=0,\qquad \qquad \qquad \qquad \qquad \qquad \qquad\qquad \text{for $x<0$},\\
\frac{1}{2\pi\mathrm{i}}\oint\limits_{\mathcal{C}_+}f_+(y,p)e^{p(x-W/2)}+\oint\limits_{\mathcal{C}_-}f_-(y,p)e^{px}\,\mathrm{d}p=\\
\qquad -\sum\limits_{k}e^{p_k(x-W/2)}\mathrm{Res}[f_+(y,p_k),\mathrm{Re}[p_k]>0]+\sum\limits_{k}e^{p_k x}\mathrm{Res}[f_-(y,p_k),\mathrm{Re}[p_k]<0], \quad \text{for $0\leq x\leq W/2$}\\
\frac{1}{2\pi\mathrm{i}}\oint\limits_{\mathcal{C}_-}f(y,p)e^{px}\,\mathrm{d}p=0, \qquad \qquad \qquad \qquad \qquad \qquad \qquad\qquad\text{for $x>W/2$}.
\end{cases}
\label{eq-Airy-full}
\end{align}
Note the different signs in the two terms in the Airy potential within the sheet domain ($x\in[0,W/2]$); "{\it a closed contour is described in a positive direction with respect to the domain enclosed by the contour if, with respect to some point inside the domain, the contour is traversed in a counterclockwise direction}"~\cite{MorseFeshbachI}.

\begin{figure}[t!]
    \centering
        \begin{tikzpicture}
\def\gap{-0.2}
\def\bigradius{3}
\def\littleradius{0.5}

\draw [help lines,->] (-1.25*\bigradius, 0) -- (1.25*\bigradius,0);
\draw [help lines,->] (0, -1.25*\bigradius) -- (0, 1.25*\bigradius);
\draw[line width=1pt,   decoration={ markings,
  mark=at position 0.2 with {\arrow[line width=1.2pt]{>}},
  mark=at position 0.4 with {\arrow[line width=1.2pt]{>}},
  mark=at position 0.7 with {\arrow[line width=1.2pt]{>}},
  mark=at position 0.9 with {\arrow[line width=1.2pt]{>}}},
  postaction={decorate}]
  let
     \n1 = {90+asin(\gap/2/\bigradius)}
  in (\n1:\bigradius) arc (\n1:-\n1:\bigradius)
  -- cycle;

\node at (3.6,-0.3){$\mathrm{Re}[p]$};
\node at (-0.6,3.53) {$\mathrm{i}\,\mathrm{Im}[p]$};
\node at (1.8,2.8) {$\mathcal{C}_+$};
\node at (-3,3.6) {\textbf{a}};

\end{tikzpicture}
    \begin{tikzpicture}
\def\gap{-0.2}
\def\bigradius{3}
\def\littleradius{0.5}

\draw [help lines,->] (-1.25*\bigradius, 0) -- (1.25*\bigradius,0);
\draw [help lines,->] (0, -1.25*\bigradius) -- (0, 1.25*\bigradius);
\draw[line width=1pt,   decoration={ markings,
  mark=at position 0.2 with {\arrow[line width=1.2pt]{>}},
  mark=at position 0.4 with {\arrow[line width=1.2pt]{>}},
  mark=at position 0.7 with {\arrow[line width=1.2pt]{>}},
  mark=at position 0.9 with {\arrow[line width=1.2pt]{>}}},
  postaction={decorate}]
  let
     \n1 = {90+asin(\gap/2/\bigradius)}
  in (\n1:\bigradius) arc (\n1:270-asin(\gap/2/\bigradius):\bigradius)
  -- cycle;

\node at (3.6,-0.3){$\mathrm{Re}[p]$};
\node at (-0.6,3.53) {$\mathrm{i}\,\mathrm{Im}[p]$};
\node at (-1.8,2.8) {$\mathcal{C}_-$};
\node at (-3,3.6) {\textbf{b}};

\end{tikzpicture}
    \caption{\textbf{Contours in the complex plane.} The closed contours over which we apply the Residue Theorem in \eqref{eq-Airy-full} --- these encircle all poles of the functions  $f_+(y,p)$ (panel \textbf{a}) and  $f_-(y,p)$ (panel \textbf{b});  the radii of circular portion of these contours must be chosen to be greater than the modulus of largest pole, $|p_k|$.}
    \label{fig:contours}
\end{figure}

It is only a matter of some (tedious) algebra to calculate the residues in~\eqref{eq-Airy-full} (using ~\eqref{eq-residues} and the corresponding definitions of the functions $f_{\pm}(y,p)$~\eqref{eq-redef-f}). Finally, we obtain the result in~\eqref{eq-Airy-final}, with
\begin{subequations}    
\begin{align}
C_k=&-\frac{\sin p_k}{\cos^2 p_k}\sum_m \frac{(p_k\alpha_m+\beta_m)}{\left[p_k^2-\left(\frac{m\pi}{2}\right)^2\right]^2}m\pi\sin\left(\frac{m\pi}{2}\right),\\
\hat C_k=&-[f^{(-)}_{p,y}(1,-p_k)-\sin^2p_k \,f_p^{(-)}(1,-p_k)]\frac{\sin p_k}{\cos^2 p_k},
\end{align}
\label{eq-coeffs-Airy}
\end{subequations} $m=2k-1$ for $k\in{\mathbb Z}^+$; here, to write things more compactly, we have introduced the function $f^{(-)}_{p}(y,p)$
\begin{align}
f^{(-)}_{p}(y,p)=f_p(y,p)+e^{-p W/2}\sum_m\frac{(p\,\alpha_m+\beta_m)}{\left[p^2-\left(\frac{m\pi}{2}\right)^2\right]^2}\cos\left(\frac{m\pi}{2}y\right).
\end{align}

\subsection*{Near-threshold analysis}
Based on the planar stress field results discussed previously, we examine the buckling of a clamped square sheet, of side $\sim W$ and thickness $t$, subject to a uniform bi-axial stress field of modulus $\sim T_{0}$, which is tensile along the $y$ axis, $\sigma_{yy}\sim T_{0}$, and compressing the $x$ axis, $\sigma_{xx}\sim-T_{0}$.

We assume then an undulating displacement field $\zeta(x,y)=\delta\cdot g_{\lambda}(x,y)$.
The amplitude $\delta$ is infinitesimal and $g_{\lambda}(x,y)$ is some undulating function along the compressive axis with a characteristic length $\lambda$, yet unknown. Along the tensile axis, $g_{\lambda}(x,y)$ simply suppresses the amplitude at the edges $\pm W/2$.
The excess bending energy associated with this perturbation is $U_{\mathrm{bend}}\sim B\int~(\nabla^2\zeta)^2\mathrm{d}A\sim B\, W^2\left[\left(\frac{\delta}{\lambda^2}\right)^2+\left(\frac{\delta}{W^2}\right)^2\right]$ --- the full sheet width at the edges is clamped, and hence going out of plane effectively means that the sheet develops a non-zero curvature along both the tensile and the compressing directions. For a given $\lambda$, the excess stretching energy of this double curvature is $U_{\mathrm{stretch}}\sim\int\sigma:\epsilon~\mathrm{d}A\sim T_{0} W^2 \delta^2/W^2-\sigma_{xx}(\lambda) W^2 \delta^2/\lambda^2$, and hence the planar state becomes unstable only if the compressive load exceeds
\begin{align}
\sigma_{xx}(\lambda)\sim \frac{B}{\lambda^2}+\left(\frac{B}{W^4}+\frac{T_{0}}{W^2}\right)\lambda^2.
\end{align} However, since $\sigma_{xx}$ is also proportional to $T_{0}$, we find that
\begin{align}
T_{0}\sim \frac{B}{\lambda^2}\frac{\left(1+\frac{\lambda^4}{W^4}\right)}{\left(1+\frac{\lambda^2}{W^2}\right)}.
\end{align} 
Since the wavelength is limited by $W$, we express $\lambda$ as $\lambda=\xi W$, with $\xi\in(0,1]$; then we can write $T_{0}\sim\frac{B}{W^2}g(\xi)$, with $g(\xi)=\frac{1}{\xi^2}\frac{(1+\xi^4)}{(1+\xi^2)}$. For $\xi\ll 1$, $g(\xi)\sim \xi^{-2}>1$; while for $\xi\sim 1$ (\emph{i.e.} $\xi=1-\varepsilon$, $\varepsilon\ll 1$), $g(\xi)\sim (1-\varepsilon)^{-2}>1$. Thus, we have that $g(\xi<1)> 1$ and $g(\xi=1)=1$; or, in other words, $g(\xi=1)=1$ is the minimum value of $g(\xi)$ in the interval $\xi\in(0,1]$. This minimum gives the threshold value for the tension
\begin{align}
T_0^c\sim \frac{B}{W^2}.
\end{align} In terms of the longitudinal in-plane end-stretching imposed, $\epsplan$, we conclude that the end-stretched sheet buckles when the longitudinal in-plane displacement $\epsplan$ exceeds $\epsplan^c$, with
\begin{align}
\epsplan^c\sim \frac{B}{YW^2}.
\end{align}

\section{Effect of vanishing Poisson's ratio}\label{sec:Zero-Poisson}

We examine the limit of vanishing Poisson's ratio $\nu\to0$ in the tension-induced transverse actuation of a sheet. In Fig.~\ref{fig:PlanarStressTransverse} (upper row) we plot the transverse component of the stress tensor~\eqref{eq-transv-stress} for a localized-loaded sheet, $\Wload\ll W$. Over a very wide range of Poisson's ratios, $0.003\leq\nu\leq0.3$, the results show that the compression zone --- depicted in blue --- persists when $\nu\to0$. We emphasize that the same color scale is used in these plots and, hence, that the magnitude of this compression is also maintained as $\nu\to0$. Conversely, when we plot this same transverse component for a full-width loaded sheet, $\Wload=W$, both the compressed area and the maximum compression shrink as $\nu\to0$ (see  Fig.~\ref{fig:PlanarStressTransverse} bottom row, for which we use Eq.~(9b) in Ref.~\cite{XinDavidovitch21}).

This comparison hints that the loading geometry is key to the localized TUG-folding demonstrated in the main text and distinguishes it from the tensional wrinkling of Cerda and Mahadevan~\cite{cerda2003geometry}. This robustness to varying $\nu$ suggests a geometric origin to the compression.
To gain insight into the mechanism that converts the longitudinal tension into a transverse compression for the different geometries, we consider a reductionist model of the sheet (see e.g.~the kinematic analysis in the model by Celli \emph{et al.}~\cite{CelliMcMahan23}). Consisting of four rigid right-angled triangles whose acute-angled vertices are connected with a hinge (see Fig.~\ref{fig:PlanarStressTransverse}, leftmost panels), the model inscribes a void rectangle that allows for `rigid' deformations when subject external loads. In this construction, a uniaxial point-force generates a torque with respect to the hinges outside the loading axis, and hence the four triangles rotate about them (see Fig.~\ref{fig:PlanarStressTransverse}, leftmost upper panel), generating transverse compression from uniaxial stretching. However, for a full-width loading the net torque vanishes, there is no rotation (Fig.~\ref{fig:PlanarStressTransverse}, leftmost lower panel) and hence no transverse compression.

Translating this heuristic model into the rectangular elastic sheet, a point loading converts into transverse compression by `rotation' of the bulk material. Thus, the compression persists despite vanishing Poisson's ratio.
By contrast, the full-width loading stretches the bulk material longitudinally, and the transverse compression is hence a Poisson effect that vanishes when $\nu\to0$.

\begin{figure}[t!]
    \centering
    \includegraphics[width=\linewidth]{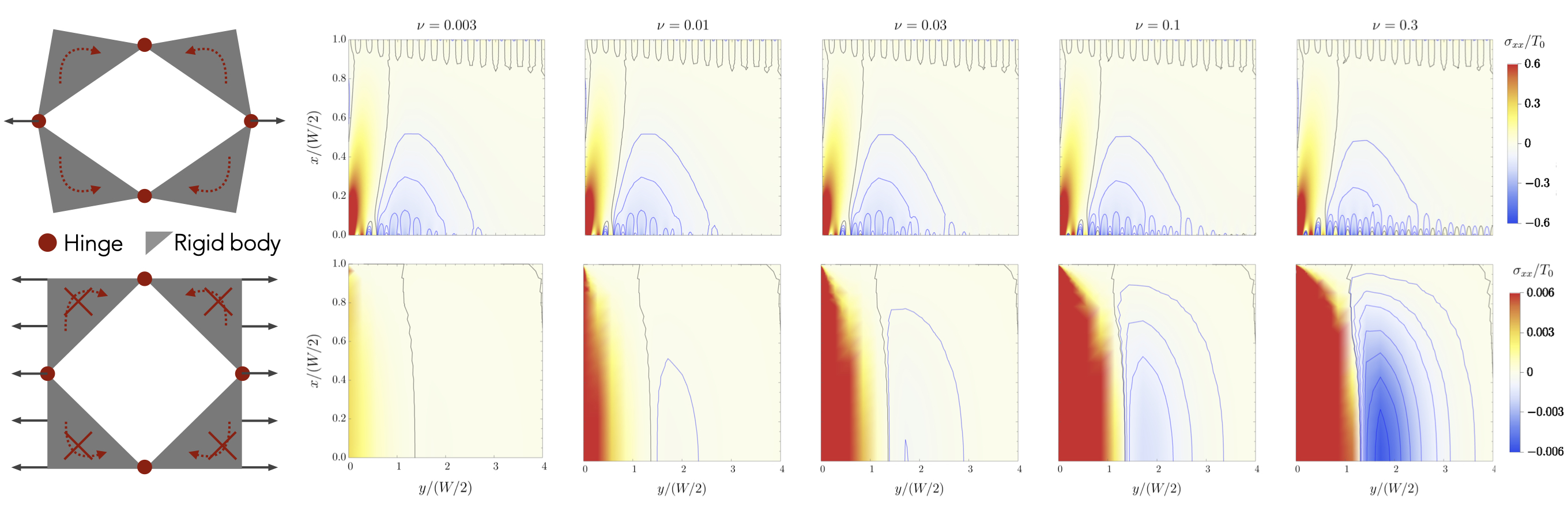}
    \caption{\textbf{Transverse stress in the planar state for different Poisson's ratios.} 
    \textbf{Leftmost panels:} Reductionist model of a longitudinally-loaded sheet consisting of four rigid right-angled triangles whose acute-angled vertices are connected with a hinge, thus forming a rectangular sheet — in absence of external loading — that inscribes a void rectangle. \textbf{Upper row:} Localized-loaded sheet. \textbf{Lower row:} Full-width loaded sheet. \textbf{From left to right:} Density plots of the transverse stress at increasing Poisson's ratio ($\nu=0.003,0.01,0.03,0.1,0.3$ as per the labels) for a localized-loaded sheet (upper row) and a full-width loaded sheet (lower row).
    Results for the localized loading are by the expression~\eqref{eq-transv-stress} (the longitudinal coordinate, $y$, is shifted $y\to y+1$). For the full-width loading, the planar stress is derived from the Airy potential Eq.~(9b) in Ref.~\cite{XinDavidovitch21} (after normalizing its coefficients by the corresponding Poisson's ratio in each case). The same colorbar (only shown in the rightmost panels of each row) applies to all panels within a row. Contour lines are displayed at $\sigma_{xx}/T_{0}=(0,-0.05,-0.1,-0.15)$ (localized-loaded sheet) and at $\sigma_{xx}/T_{0}=(0, -0.0001, -0.0005, -0.001, -0.002,-0.003, -0.004,-0.005)$ (full-width loaded sheet). The aspect ratio is $W/L=1/4$ and we sum $n=32$ terms in the Airy potential's series to derive the transverse stress in each panel.
    }
    \label{fig:PlanarStressTransverse}
\end{figure}

\section{Wrinkling at larger strain}

The analysis of the main text showed that a sheet subject to a localized `tug' force buckles and folds at smaller strains than does a sheet clamped over its full-width does. However, as the strain increases the width of the  doubly curved region, denoted $\omega$, shrinks (see Eq.~\eqref{eqn:CurvedWidth} of the main text) and so must, at some strain, become comparable to the loading width $\Wload$. At this point we expect the situation to change since, within this doubly-curved region at least, the loading is no longer `point-like'. In particular,  in this region, and for large strains, the problem  becomes analogous to the CM problem~\cite{cerda2003geometry} and wrinkles may be expected  to form.

\begin{figure}[htb]
    \centering
    \includegraphics[width=0.7\linewidth]{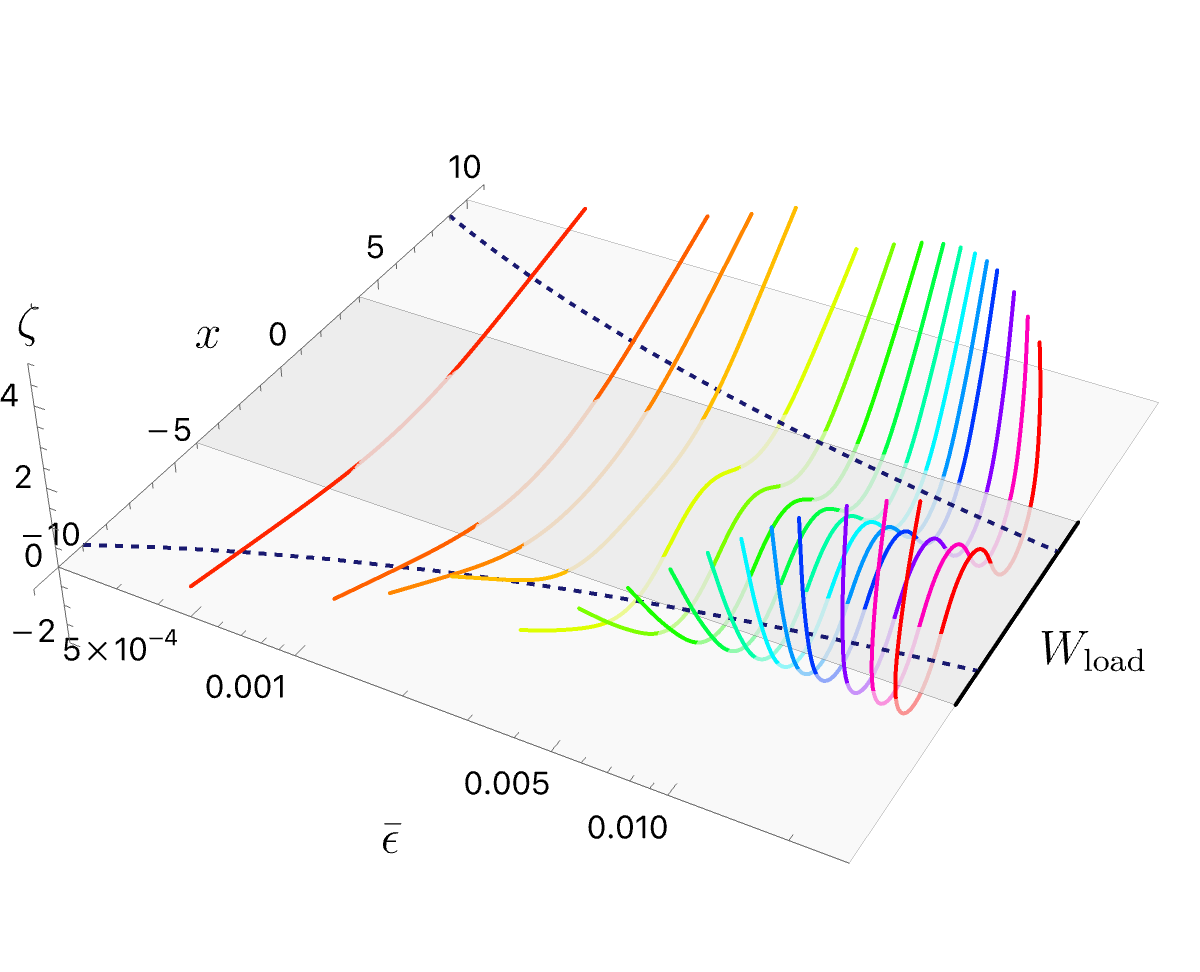}
    \caption{\textbf{Wrinkling at larger strain.} Cross-sectional profiles $\zeta(x,y=0)$, obtained from FEM simulations, of a sheet   subject to different longitudinal strains  exerted by pulling clamps of width $\Wload/W=0.4$. The profiles  at different imposed strains are shown by solid curves, with color used to distinguish profiles at different strains. Here, $W=20, L/W=3,t/W=1\times 10^{-3}$   and the load width is emphasized by the  gray shaded region. We note that the localized TUG folding studied here gives way to wrinkles similar to those in the CM problem when the strain $\epsbar$ becomes large enough: we use Eq.~(5) of Ref.~\cite{cerda2003geometry} to determine the expected wrinkle wavelength $\lambda_{\rm CM}$ and then plot $x=\pm0.5\times\lambda_{\mathrm{CM}}(\epsbar)$  (dashed curves). Note that the transition to wrinkling occurs when $\Wload\approx\lambda_{\rm CM}$.}
    \label{fig:wrinkles}
\end{figure}

 We demonstrate this behavior using FEM simulations of a rectangular sheet subject to increasing tensile strains with $\Wload/W=0.4$, see Fig.~\ref{fig:wrinkles}. Notably, higher modes in the out-of-plane displacement emerge when $\Wload\simeq \lambda_{\mathrm{CM}}$, where 
 $$\lambda_{\mathrm{CM}}=2\sqrt{\pi}\left(\frac{B L^2}{Y\epsbar}\right)^{1/4}$$ is the wavelength in the CM problem, Eq.~(5) of Ref.~\cite{cerda2003geometry}, and we have used $\bar{T}=Y\epsbar$ to eliminate tension. A lower (asymmetric) mode would in principle be stable at smaller strains, i.e. when $\Wload\ < \lambda_{\mathrm{CM}}$, yet we do not observe this in the simulations because we consider perfectly symmetric sheets—a small imperfection is only introduced to break the up-down symmetry.

\vspace{2cm}